\documentclass[11pt]{article}
\usepackage{amsmath}
\usepackage{amssymb}
\usepackage{amsfonts}
\usepackage{graphicx}
\usepackage{float}

\allowdisplaybreaks
\input{tcilatex}

\begin{document}

\begin{center}
{\Large Aircraft navigation based on differentiation-integration observer}

Xinhua Wang$^{1}$, Lilong Cai$^{2}$

$^{1}${\small Department of Electrical and Electronic Engineering,
University of Nottingham,}

{\small University Park, Nottingham, NG7 2RD, United Kingdom (Email:
xinhua.wang1@nottingham.ac.uk)}

$^{2}${\small Department of Mechanical and Aerospace Engineering,}

{\small Hong Kong University of Science and Technology, Hong Kong, China
(Email: melcai@ust.hk)}
\end{center}

\emph{Abstract:} In this paper, a generalized differentiation-integration
observer is presented based on sensors selection. The proposed
differentiation-integration observer can estimate the multiple integrals and
high-order derivatives of a signal, synchronously. The parameters selection
rules are presented for the differentiation-integration observer. The
theoretical results are confirmed by the frequency-domain analysis. The
effectiveness of the proposed observer are verified through the numerical
simulations on a quadrotor aircraft: i) through the
differentiation-integration observer, the attitude angle and the
uncertainties in attitude dynamics are estimated synchronously from the
measurements of angular velocity; ii) a control law is designed based on the
observers to drive the aircraft to track a reference trajectory.


\section{Introduction}

\markboth{}
{Murray and Balemi: Using the Document Class IEEEtran.cls}%
\setcounter{page}{1}Integration and differentiation are important components
in almost all industrial applications. Their problems are of estimating the
values $I\left( a\right) =\int_{0}^{t}\cdots \int_{0}^{s}a\left( \sigma
\right) d\sigma \cdots d\tau $ and $D_{i}\left( a\right) =\frac{d^{i}a\left(
t\right) }{dt^{i}}$. The positions, velocities and accelerations are the
important elements for many systems. In an inertial navigation system (INS),
the inertial measurement unit (IMU) typically measures the three-axial
angular velocity and the three-axial linear acceleration, respectively. To
obtain the attitude angle and angular acceleration of the device, the
angular velocity signals are integrated and differentiated, respectively.
For a long-time navigation, the drift phenomenon of INS is mainly brought
out by the usual integration methods. They cannot restrain the effect of
stochastic noise (especially non-zero mean noise). The noise leads to the
accumulation of additional drift in the integrated signal.

The algorithms of differentiation and integration have been studied by a
number of researchers [1]-[23]. The linear high-gain differentiators [2, 3]
can provide the estimations of signal derivatives. In another study, a
differentiator via high-order sliding modes algorithm was proposed [4, 5].
In [6]-[9], the continuous nonlinear differentiators based on finite-time
stability were presented to provide the smooth estimations of signal
derivatives. However, the differentiators did not consider the signal
integral estimations [1]-[10].

Some numerical methods were proposed to estimate signal integral [11]: i)
The trapezoidal rule; ii) Simpson's rule. For the above numerical
integrating methods, if stochastic non-zero mean noise exists in signal,
then such noise leads to the accumulation of additional drift in the
integrated signal. As we known, the desired integral operators $1/s$ and $%
1/s^{2}$ are irrational, and they cannot be calculated directly. Some
approximated methods were presented to estimate signal integral: IIR digital
integrator [12, 13], the Newton-Cotes digital integrators [14], precision
digital integrator [15], non-inverting integrator [16], the developed
infinite impulse response digital integrators [17, 18], the low-frequency
differential differentiators [19, 20]. However, for the aforementioned
integrators [12]-[20], only onefold integral was calculated, and the
synchronous estimations of derivatives and integrals were not considered.
Some integrators were implemented using the hardware units, where the
circumstances usually affect the parameters, for instance, the temperature
in the circuit changes. Thus, the estimation precisions are affected
adversely. Moreover, they are easily infected by stochastic noise, and the
drift phenomena are inevitable in such systems. In order to reduce the
noise, additional filters must be added. In [21] and [22], a
fractional-order integrator has been presented, and a rational transfer
function was proposed to approximate the irrational integrator $1/s^{m}$.
However, the limitation of $0<m<1$ limits its application. The onefold and
double integrals are necessary in many navigation systems. The Kalman filter
can estimate position and velocity from the acceleration measurement [23].
However, it is supposed that the process noise covariance and measurement
noise covariance are required to be zero-mean Gaussian distributed, and the
process noise covariance is uncorrelated to the estimation error. These
assumptions are different from the practical noise in signal. The inaccurate
noise requirements may lead to the estimate drifts of position and velocity.

In [24], a nonlinear double-integral observer was presented to estimate
synchronously the onefold and double integrals of a signal, and a
generalized multiple integrator was designed to estimate the multiple
integrals [25]. In [26], a nonlinear integral-derivative observer was
proposed to estimate synchronously the integral and derivative of a signal.
The parameters selection is required to be satisfied with Routh-Hurwitz
Stability Criterion and the iterative equation relations. Moreover, the
nonlinear observers in [26] are complicated and difficult to compute.The
existing hardware computational circumstances affect the nonlinear function
implementations adversely, i.e., the implementation of the these nonlinear
observers in many digital processors is difficult. Due to the existence of
such many parameters, the parameters regulation of this nonlinear observer
is complicated.

In this paper, a generalized high-order linear differentiation-integration
observer is presented, which can estimate the multiple integrals and
high-order derivatives of a signal, synchronously. Different from the
nonlinear observer theories in [26], the classical theory of linear system
can be used to prove its stability, and the Bode plots are adopted to
analyze its robustness. The parameters selection become relaxed, and it is
only required to be satisfied with a simple Hurwitz condition. The existing
layout of perturbation parameters in singular perturbation technique [27,
28] is only suitable to estimate the derivatives of a signal. In this
differentiation-integration observer, a new distribution of perturbation
parameters is presented for the requirement of synchronous estimation of the
integrals and derivatives. The parameters selection rules and robustness
analysis for the differentiation-integration observer are presented based on
frequency-domain analysis.

Finally, we use the mathematical model and reference trajectory of the
quadrotor aircraft described in [29], and the computational analysis and
simulation are presented to observe the performances of the proposed
observer. Usually controlling a quadrotor aircraft to track a reference
trajectory needs the information of the position and attitude. Quadrotor
aircraft control has been an active area of investigation for several years
[30]-[34]. However, these strategies are dependent on the accurate model,
and all the states are required to be known. In [29], although the
uncertainies were considered, the attitude angle was supposed to be known.
For the system of a quadrotor aircraft, we consider that the information of
flying velocity and attitude angle is not provided. Moreover, quadrotor
aircrafts are underactuated mechanical systems, and aerodynamic disturbance,
unmodelled dynamics and parametric uncertainties are not avoidable in
modeling. Based on the presented differentiation-integration observer, this
paper provides a tracking control method for a quadrotor aircraft by using
the measurements of position and angular velocity. The unknown velocity,
attitude angle and uncertainties are reconstructed by the observers.
Furthermore, a controller is designed to stabilize the flight dynamics.

\section{Generalized differentiation-integration observer}

\emph{2.1 Configuration of differentiation-integration observer}

The goal of the differentiation-integration observer design is to estimate
the needed states from the different sensors, for instance:

\emph{1) Estimation of velocity and acceleration from position:} In a GPS,
position $p(t)$ is known, we want to obtain the velocity and acceleration.
Therefore, we expect to design an observer to estimate the first-order
derivative $\dot{p}\left( t\right) $ and second-order derivative $\ddot{p}%
\left( t\right) $ of the signal $p(t)$, synchronously. The observer should
include the state $x=(x_{1},x_{2},x_{3})$, and the configuration of the
observer is described as follow:

\begin{eqnarray}
\dot{x}_{1} &=&x_{2};\dot{x}_{2}=x_{3};  \notag \\
\dot{x}_{3} &=&f(x_{1}-p(t),x_{2},x_{3})
\end{eqnarray}%
where the first state $x_{1}$ of observer (1) points to the input signal $%
p(t)$. If the state $x_{1}\ $estimates $p(t)$, and the system is stable,
then, from the integral-chain relations of $x_{1},x_{2},x_{3}$ in system
(1), the first-order and second-order derivatives of signal $p(t)$ can be
estimated, synchronously.

\emph{2) Estimation of attitude angle and angular acceleration from angular
velocity:} In an IMU, angular velocity $\omega (t)$ can be measured
directly. We want to obtain the attitude angle and angular acceleration.
Therefore, we expect to design an observer to estimate the integral $%
\int_{0}^{t}\omega \left( \sigma \right) d\sigma $ and the derivative $\dot{%
\omega}\left( t\right) $ of signal $\omega (t)$, synchronously. The observer
should include the state $x=(x_{1},x_{2},x_{3})$, and the configuration of
the observer is shown as follow:

\begin{eqnarray}
\dot{x}_{1} &=&x_{2};\dot{x}_{2}=x_{3};  \notag \\
\dot{x}_{3} &=&f(x_{1},x_{2}-a(t),x_{3})
\end{eqnarray}%
where the second state $x_{2}$ of observer (2) points to the input signal $%
\omega (t)$. If the state $x_{2}\ $estimates $\omega (t)$, and the system is
stable, then, from the integral-chain relations of $x_{1},x_{2},x_{3}$ in
system (2), the onefold integral and first-order derivative of signal $%
\omega (t)$ can be estimated, synchronously.

\emph{3) Estimation of position and velocity from acceleration:} In the
third case, the acceleration $a_{c}(t)$ is measured by a accelerometer, and
we want to obtain the position and the velocity. Therefore, we expect to
design an observer to estimate the onefold integral $\int_{0}^{t}a_{c}\left(
\sigma \right) d\sigma $ and the double integral $\int_{0}^{t}%
\int_{0}^{s}a_{c}\left( \sigma \right) d\sigma d\tau $ of signal $a_{c}(t)$,
synchronously. The observer should include the state $x=(x_{1},x_{2},x_{3})$%
, and the configuration of the observer is described as follow:

\begin{eqnarray}
\dot{x}_{1} &=&x_{2};\dot{x}_{2}=x_{3};  \notag \\
\dot{x}_{3} &=&f(x_{1},x_{2},x_{3}-a(t))
\end{eqnarray}%
where the third state $x_{3}$ of observer (3) points to signal $a_{c}(t)$.
If the state $x_{3}\ $estimates $a_{c}(t)$, and the system is stable, then,
from the integral-chain relations of $x_{1},x_{2},x_{3}$ in system (3), the
onefold and double integrals of signal $a_{c}(t)$ can be estimated,
synchronously.

\emph{4) Generalized cases:} Generally, for signal $a(t)$ measured from a
sensor, we expect to design an observer to estimate the multiple integrals
up to ($p-1$)th multiple and high-order derivatives up to ($n-p$)th order,
synchronously, where $p\in \{1,\cdots ,n\}$ corresponds to different
sensors. Let the $i$th-multiple integral of signal $a(t)$ be $I_{p-i}\left(
t\right) =\underset{i}{\underbrace{\int_{0}^{t}\cdots \int_{0}^{s}}}a\left(
\sigma \right) \underset{i}{\underbrace{d\sigma \cdots d\tau }}$, where $%
i\in \{1,\cdots ,p-1\}$; $I_{p}\left( t\right) =a(t)$; and the ($r-p$%
)th-order derivative of signal $a(t)$ be $I_{r}\left( t\right) =a^{\left(
r-p\right) }\left( t\right) $, $r=p+1,\cdots ,n$. The observer should
include the state $x=(x_{1},\cdots ,x_{p},\cdots ,x_{n})$, and the
configuration of the observer is presented as follow:

\begin{eqnarray}
\dot{x}_{1} &=&x_{2}  \notag \\
&&\cdots  \notag \\
\dot{x}_{p} &=&x_{p+1}  \notag \\
&&\cdots  \notag \\
\dot{x}_{n} &=&f(x_{1},\cdots ,x_{p}-a(t),\cdots ,x_{n})
\end{eqnarray}%
where the $p$-th state $x_{p}$ of observer (4) points to signal $a(t)$. If
the state $x_{p}\ $estimates $a(t)$, and the system is stable, then, from
the integral-chain relations of $x_{1},\cdots ,x_{p},\cdots ,x_{n}$ in
system (4), the multiple integrals and high-order derivatives of signal $%
a(t) $ can be estimated, synchronously.

The goals of the observation are: 1) synchronous estimation of multiple
integrals and high-order derivatives of a signal; 2) regulation of low-pass
frequency bandwidth through the ease of parameter selection, and sufficient
high-frequency noise rejection.

\bigskip

\emph{2.2 Existence conditions of Hurwitz characteristic polynomial}

Before constructing the explicit form of the differentiation-integration
observer (4), we propose a $n$th-order characteristic polynomial

\begin{equation}
s^{n}+k_{n}s^{n-1}+\cdots +\frac{k_{p}}{\varepsilon ^{p-c(p)}}s^{p-1}+\cdots
+k_{2}s+k_{1}
\end{equation}%
where, $p\in \{1,\cdots ,n\}$, and

\begin{equation}
c(p)=\left\{
\begin{array}{c}
1,p=1 \\
0,p>1%
\end{array}%
\right.
\end{equation}%
Here, the Hurwitz conditions of the polynomial (5) is considered. The
polynomial is used for construct a linear differentiation-integration
observer. We consider the following question: For the arbitrary $\varepsilon
\in \left( 0,1\right) $, by selecting parameters $k_{1}$, $\cdots $, $k_{n}$%
, whether can all the positive integers $n$ and $p\in \{1,\cdots ,n\}$ make
the polynomial (5) Hurwitz?

For instance, we can find that: 1) for the arbitrary $\varepsilon \in \left(
0,1\right) $, when we select $n=4$ and $p=2$, the polynomial (5) cannot be
Hurwitz; 2) for the arbitrary $\varepsilon \in \left( 0,1\right) $, when $%
n=5 $ and $p\in \{2,3,4,5\}$, the polynomial (5) cannot be Hurwitz.

\bigskip

For the arbitrary $\varepsilon \in \left( 0,1\right) $, the following lemma
presents the selections of $n$ and $p$ to make polynomial (5) Hurwitz.

\emph{Lemma 1: }For the arbitrary $\varepsilon \in \left( 0,1\right) $, in
the following cases, the polynomial (5) can be Hurwitz:

a) $n\in \{1,2,\cdots \}$ and $p=1$: $k_{i}>0$ (where $i=1,\cdots ,n$) are
selected such that the polynomial $s^{n}+\sum\limits_{{i=1}}^{{n}%
}k_{i}s^{i-1}$ is Hurwitz.

b) $n=2$ and $p=2$: $k_{1}>0,k_{2}>0$.

c) $n=3$ and $p\in \left\{ 2,3\right\} $: when $p=2$, $k_{1}>0,k_{3}>0$ and $%
k_{2}>\varepsilon ^{2}\frac{k_{1}}{k_{3}}$; when $p=3$, $k_{1}>0,k_{3}>0$
and $k_{2}>\varepsilon ^{3}\frac{k_{1}}{k_{3}}$.

d) $n=4$ and $p=3$: $k_{1}>0,k_{4}>0,k_{3}>\varepsilon ^{3}\frac{k_{2}}{k_{4}%
}$ and $k_{2}>\varepsilon ^{3}\frac{k_{4}^{2}k_{1}+k_{2}^{2}}{k_{4}k_{3}}$.

The proof of Lemma 1 is presented in Appendix.

\bigskip

\emph{2.3 Design of generalized differentiation-integration observer}

In the following, the singular perturbation technique will be used to design
a generalized differentiation-integration observer, and Theorem 1 is
presented as follow.

\emph{Theorem 1:} For system

\begin{eqnarray}
\dot{x}_{i} &=&x_{i+1};i=1,\cdots ,n-1  \notag \\
\varepsilon ^{n+1-c(p)}\dot{x}_{n} &=&-\sum\limits_{{i=1,i\neq p}}^{{n}%
}k_{i}\varepsilon ^{i-c(p)}x_{i}-k_{p}\left( x_{p}-a\left( t\right) \right)
\end{eqnarray}%
where, $p\in \left\{ 1,\cdots ,n\right\} $, and

\begin{equation}
c(p)=\left\{
\begin{array}{c}
1,p=1 \\
0,p>1%
\end{array}%
\right.
\end{equation}%
$a\left( t\right) $ is the signal that can be directly measured, and it\ is
continuous, integrable and ($n-p+1$)th-order derivable. Let $a_{p-i}\left(
t\right) =\underset{i}{\underbrace{\int_{0}^{t}\cdots \int_{0}^{s}}}a\left(
\sigma \right) \underset{i}{\underbrace{d\sigma \cdots d\tau }}$, $i\in
\{1,\cdots ,p-1\}$; $a_{p}\left( t\right) =a\left( t\right) $; $a_{r}\left(
t\right) =a^{\left( r-p\right) }\left( t\right) $, $r=p+1,\cdots ,n$; $%
\varepsilon \in \left( 0,1\right) $ is the perturbation parameter; $%
k_{1},\cdots ,\frac{k_{p}}{\varepsilon ^{p-c(p)}},\cdots ,k_{n}>0$ are
selected such that $s^{n}+k_{n}s^{n-1}+\cdots +\frac{k_{p}}{\varepsilon
^{p-c(p)}}s^{p-1}+\cdots +k_{2}s+k_{1}$ is Hurwitz, then the following
conclusions hold:

\begin{equation}
\underset{\varepsilon \rightarrow 0}{\lim }x_{i}=a_{i}\left( t\right) ,\text{%
for }i\in \{1,\cdots ,n\}
\end{equation}%
and observer (7) is stable.

\emph{Proof:} The Laplace transformation of (7) can be calculated as follow:

\begin{eqnarray}
sX_{i}\left( s\right) &=&X_{i+1}\left( s\right) ;i=1,\cdots ,n-1  \notag \\
\varepsilon ^{n+1-c(p)}sX_{n}\left( s\right) &=&-\sum\limits_{{i=1,i\neq p}%
}^{{n}}k_{i}\varepsilon ^{i-c(p)}X_{i}\left( s\right) -k_{p}\left(
X_{p}\left( s\right) -A\left( s\right) \right)
\end{eqnarray}%
where $X_{i}\left( s\right) $ and $A\left( s\right) $ denote the Laplace
transformations of $x_{i}$ and $a\left( t\right) $, respectively, and $s$
denotes Laplace operator. From (10), we obtain

\begin{equation}
X_{i}\left( s\right) =\frac{X_{j}\left( s\right) }{s^{j-i}},i=1,\cdots
,n,j\in \left\{ 1,\cdots ,n\right\}
\end{equation}%
Therefore, Eq. (10) can be written as

\begin{equation}
s^{n-j+1}\varepsilon ^{n+1-c(p)}X_{j}\left( s\right) =-\sum\limits_{{%
i=1,i\neq p}}^{{n}}k_{i}\varepsilon ^{i-c(p)}\frac{X_{j}\left( s\right) }{%
s^{j-i}}-k_{p}\left( \frac{X_{j}\left( s\right) }{s^{j-p}}-A\left( s\right)
\right)
\end{equation}%
Then, it follows that

\begin{equation}
\frac{X_{j}\left( s\right) }{A\left( s\right) }=\frac{k_{p}}{%
s^{n-j+1}\varepsilon ^{n+1-c(p)}+\sum\limits_{{i=1,i\neq p}}^{{n}}\frac{%
k_{i}\varepsilon ^{i-c(p)}}{s^{j-i}}+\frac{k_{p}}{s^{j-p}}}
\end{equation}%
i.e.,

\begin{equation}
\frac{X_{j}\left( s\right) }{A\left( s\right) }=\frac{s^{j-1}k_{p}}{%
s^{n}\varepsilon ^{n+1-c(p)}+\sum\limits_{{i=1,i\neq p}}^{{n}%
}s^{i-1}k_{i}\varepsilon ^{i-c(p)}+s^{p-1}k_{p}}
\end{equation}%
Therefore, we obtain

\begin{equation}
\underset{\varepsilon \rightarrow 0}{\lim }\frac{X_{j}\left( s\right) }{%
A\left( s\right) }=s^{j-p}
\end{equation}%
where $j\in \left\{ 1,\cdots ,n\right\} $ and $p\in \left\{ 1,\cdots
,n\right\} $. It means that the state $x_{i}$ approximates $a_{i}\left(
t\right) $, for $1\leq i\leq n$.

From Eq. (14), the characteristic polynomial of observer (7) is

\begin{equation}
s^{n}+\sum\limits_{{i=1,i\neq p}}^{{n}}\frac{k_{i}}{\varepsilon ^{n-i+1}}%
s^{i-1}+\frac{k_{p}/\varepsilon ^{p-c(p)}}{\varepsilon ^{n-p+1}}s^{p-1}
\end{equation}

Importantly, in order to make the system stable, the characteristic
polynomial is required to be Hurwitz. It is equivalent that

\begin{equation}
s^{n}+\sum\limits_{{i=1,i\neq p}}^{{n}}k_{i}s^{i-1}+\frac{k_{p}}{\varepsilon
^{p-c(p)}}s^{p-1}
\end{equation}%
should be Hurwitz. This concludes the proof. $\blacksquare $

From Theorem 1, we find that: the state $x_{p}$ estimates the signal $%
a\left( t\right) $, $x_{i}$ estimates the $(p-i$)th-multiple integral of
signal $a(t)$, where $i\in \{1,\cdots ,p-1\}$; $x_{r}$ estimates the the ($%
r-p$)th-order derivative of signal $a(t)$, where $r=p+1,\cdots ,n$.

\bigskip

\emph{2.4 Explicit forms of differentiation-integration observers}

From Theorem 1 and Lemma 1, the explicit forms of generalized
differentiation-integration observer can be deduced, and a corollary is
presented as follow. It includes: high-order differentiator, onefold
integrator, differentiation-integration observer, double integrator,
differentiation and double-integration observer.

\bigskip

\emph{Corollary 1: }The following differentiation-integration observers
exist:

\emph{i) High-order differentiator [35] (where }$n\in \{1,2,\cdots \}$\emph{%
\ and }$p=1$\emph{):}

\begin{eqnarray}
\dot{x}_{i} &=&x_{i+1};i=1,\cdots ,n-1  \notag \\
\varepsilon ^{n}\dot{x}_{n} &=&-k_{1}\left( x_{1}-a\left( t\right) \right)
-\sum\limits_{{i=2}}^{{n}}k_{i}\varepsilon ^{i-1}x_{i}
\end{eqnarray}%
where, $\varepsilon \in \left( 0,1\right) $; $k_{i}>0$ ($i=1,\cdots ,n$) are
selected such that the polynomial $s^{n}+\sum\limits_{{i=1}}^{{n}%
}k_{i}s^{i-1}$ is Hurwitz. For differentiator (18), the following
conclusions hold:

\begin{equation}
\underset{\varepsilon \rightarrow 0}{\lim }x_{i}=a^{(i-1)}\left( t\right) ,%
\text{for }i\in \{1,\cdots ,n\}
\end{equation}%
It can estimate the derivatives of signal $a\left( t\right) $ up to ($n-1$%
)th order.

\bigskip

\emph{ii) Onefold integrator(where }$n=2$ and $p=2$\emph{):}

\begin{eqnarray}
\dot{x}_{1} &=&x_{2}  \notag \\
\varepsilon ^{3}\dot{x}_{2} &=&-k_{1}\varepsilon x_{1}-k_{2}\left(
x_{2}-a\left( t\right) \right)
\end{eqnarray}%
where, $\varepsilon \in \left( 0,1\right) $; $k_{1}>0,k_{2}>0$. The
following conclusions hold:

\begin{equation}
\underset{\varepsilon \rightarrow 0}{\lim }x_{1}=\int_{0}^{t}a\left(
t\right) d\tau ,\underset{\varepsilon \rightarrow 0}{\lim }x_{2}=a\left(
t\right)
\end{equation}%
It can estimate the onefold integral of signal $a\left( t\right) $.

\bigskip

\emph{iii) Differentiation-integration observer (where }$n=3$\emph{\ and }$%
p=2$\emph{):}

\begin{eqnarray}
\dot{x}_{1} &=&x_{2}  \notag \\
\dot{x}_{2} &=&x_{3}  \notag \\
\varepsilon ^{4}\dot{x}_{3} &=&-k_{1}\varepsilon x_{1}-k_{2}\left(
x_{2}-a\left( t\right) \right) -k_{3}\varepsilon ^{3}x_{3}
\end{eqnarray}%
where, $\varepsilon \in \left( 0,1\right) $; $k_{1}>0,k_{3}>0$ and $%
k_{2}>\varepsilon ^{2}\frac{k_{1}}{k_{3}}$. The following conclusions hold:

\begin{equation}
\underset{\varepsilon \rightarrow 0}{\lim }x_{1}=\int_{0}^{t}a\left(
t\right) d\tau ,\underset{\varepsilon \rightarrow 0}{\lim }x_{2}=a\left(
t\right) ,\underset{\varepsilon \rightarrow 0}{\lim }x_{3}=\dot{a}\left(
t\right)
\end{equation}%
It can estimate the onefold integral and the first-order derivative of
signal $a\left( t\right) $, respectively.

\bigskip

\emph{iv) Double integrator (where }$n=3$\emph{\ and }$p=3$\emph{):}

\begin{eqnarray}
\dot{x}_{1} &=&x_{2}  \notag \\
\dot{x}_{2} &=&x_{3}  \notag \\
\varepsilon ^{4}\dot{x}_{3} &=&-k_{1}\varepsilon x_{1}-k_{2}\varepsilon
^{2}x_{2}-k_{3}\left( x_{3}-a\left( t\right) \right)
\end{eqnarray}
where, $\varepsilon \in \left( 0,1\right) $; $k_{1}>0,k_{3}>0$ and $%
k_{2}>\varepsilon ^{3}\frac{k_{1}}{k_{3}}$. The following conclusions hold:

\begin{equation}
\underset{\varepsilon \rightarrow 0}{\lim }x_{1}=\int_{0}^{t}\int_{0}^{\tau
}a\left( s\right) dsd\tau ,\underset{\varepsilon \rightarrow 0}{\lim }%
x_{2}=\int_{0}^{t}a\left( \tau \right) ,\underset{\varepsilon \rightarrow 0}{%
\lim }x_{3}=a\left( t\right)
\end{equation}%
It can estimate the onefold and double integrals of signal $a\left( t\right)
$, respectively.

\bigskip

\emph{v) Differentiation and double-integration observer (where }$n=4$\emph{%
\ and }$p=3$\emph{)}

\begin{eqnarray}
\dot{x}_{1} &=&x_{2}  \notag \\
\dot{x}_{2} &=&x_{3}  \notag \\
\dot{x}_{3} &=&x_{4}  \notag \\
\varepsilon ^{5}\dot{x}_{4} &=&-k_{1}\varepsilon x_{1}-k_{2}\varepsilon
^{2}x_{3}-k_{3}\left( x_{3}-a\left( t\right) \right) -k_{4}\varepsilon
^{4}x_{4}
\end{eqnarray}%
where, $\varepsilon \in \left( 0,1\right) $; $k_{1}>0,k_{4}>0,k_{3}>%
\varepsilon ^{3}\frac{k_{2}}{k_{4}}$ and $k_{2}>\varepsilon ^{3}\frac{%
k_{4}^{2}k_{1}+k_{2}^{2}}{k_{4}k_{3}}$. The following conclusions hold:

\begin{equation}
\underset{\varepsilon \rightarrow 0}{\lim }x_{1}=\int_{0}^{t}\int_{0}^{\tau
}a\left( s\right) dsd\tau ,\underset{\varepsilon \rightarrow 0}{\lim }%
x_{2}=\int_{0}^{t}a\left( \tau \right) ,\underset{\varepsilon \rightarrow 0}{%
\lim }x_{3}=a\left( t\right) ,\underset{\varepsilon \rightarrow 0}{\lim }%
x_{4}=\dot{a}\left( t\right) ,
\end{equation}%
It can estimate the onefold, double integrals and the first-order derivative
of signal $a\left( t\right) $, respectively.

\section{Frequency analysis and parameters selection}

In a practical problem, high-frequency noises exist in the measurement
signal. The following analysis concerns the robustness behavior of the
presented observers under high-frequency noise. We will adopt the Bode plots
to analyze the frequency characteristics of the proposed
differentiation-integration observers. Bode plots method is an indispensable
component of the bag of tools of practicing control engineers. By the
frequency analysis method, we can find that the presented
differentiation-integration observers lead to perform rejection of
high-frequency noise.

\begin{figure}[H]
\begin{center}
\includegraphics[width=5.0in]{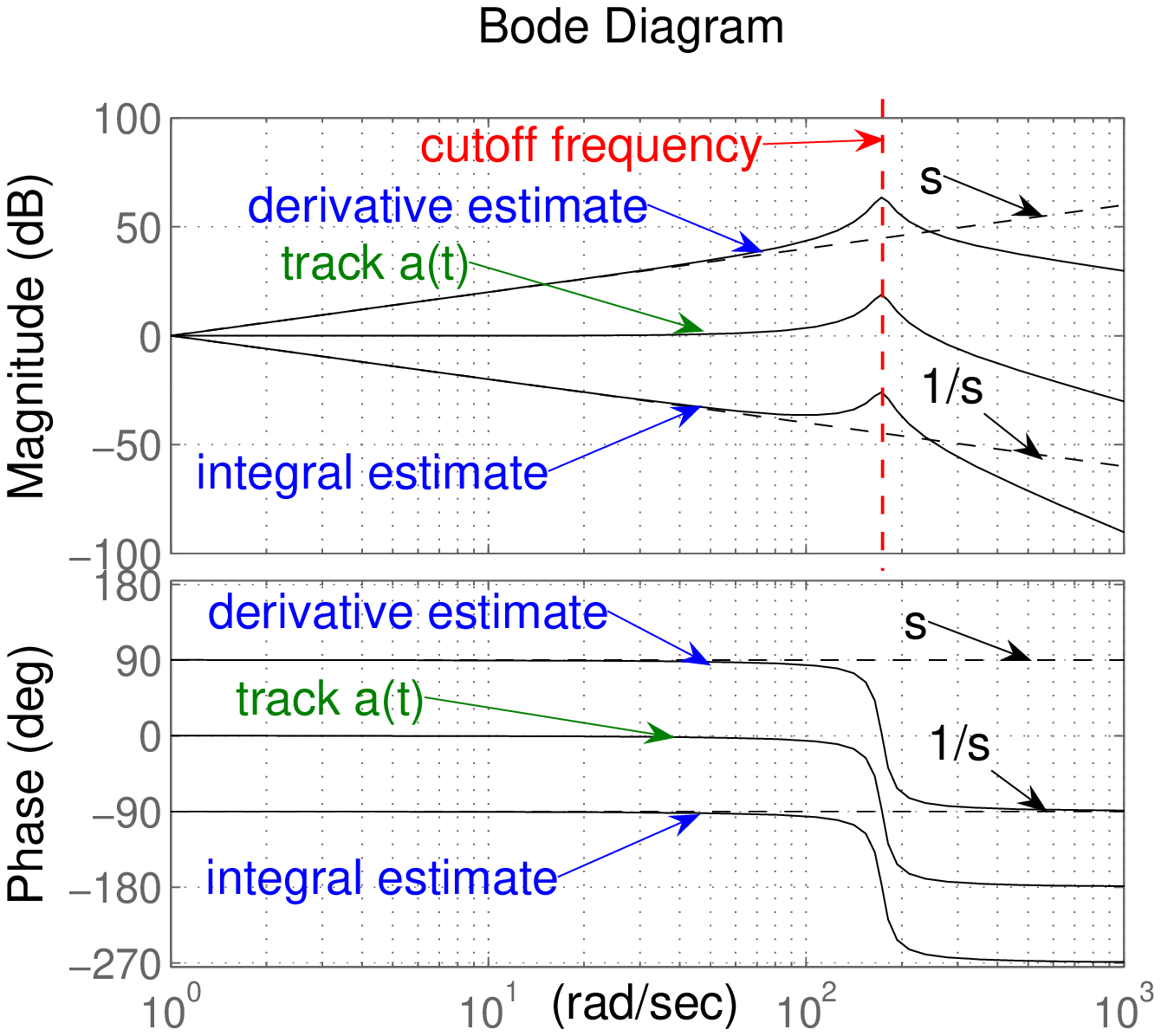}\\[0pt]
{\small 1(a) Bode plot when }$\varepsilon =0.1$\\[0pt]
\includegraphics[width=5.0in]{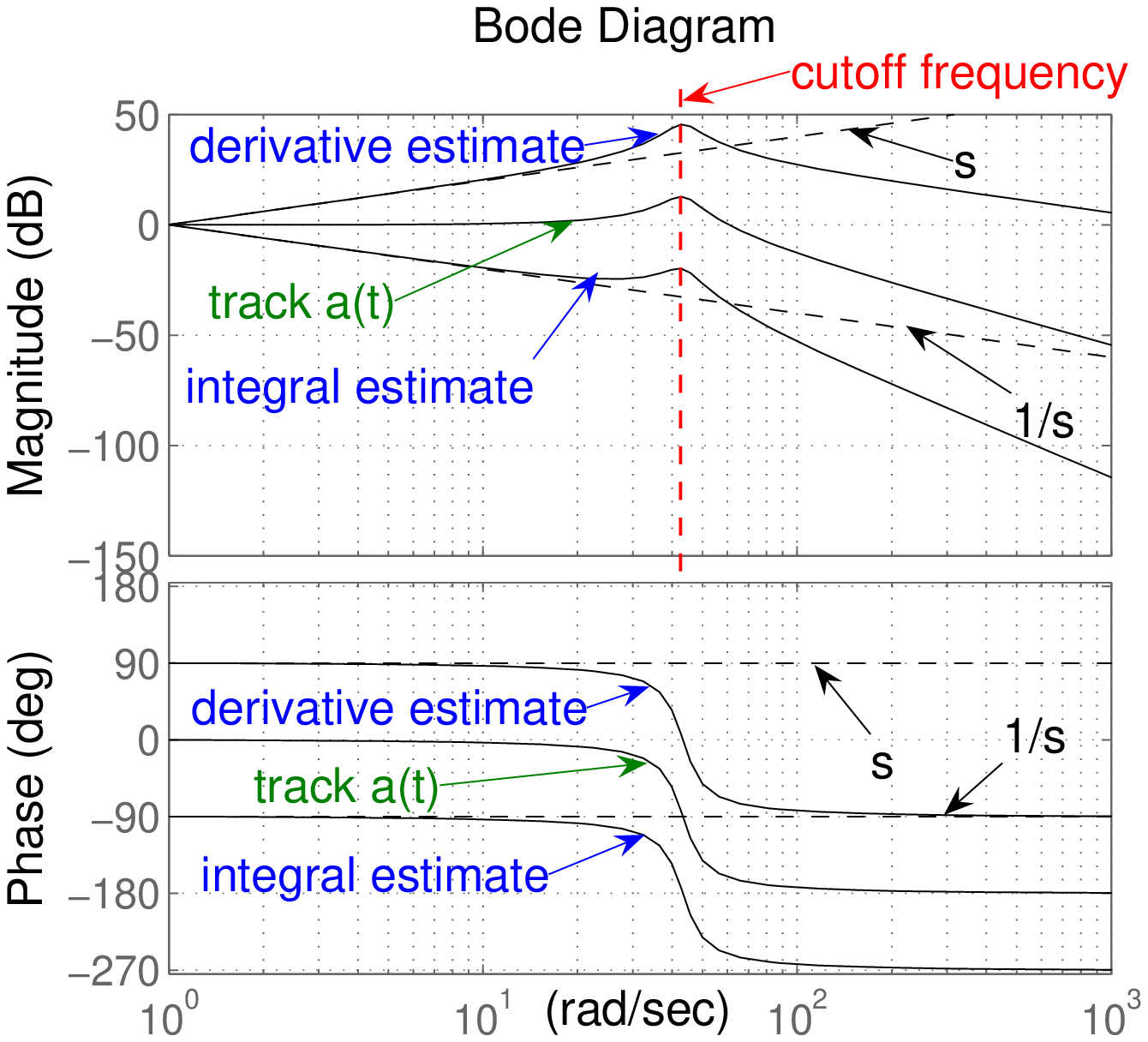}\\[0pt]
{\small 1(b) Bode plot when }$\varepsilon =0.2${\\[0pt]
{\small Figure 1 Bode plot of differentiation-integration observer}}
\end{center}
\end{figure}

\begin{figure}[H]
\begin{center}
\includegraphics[width=5.0in]{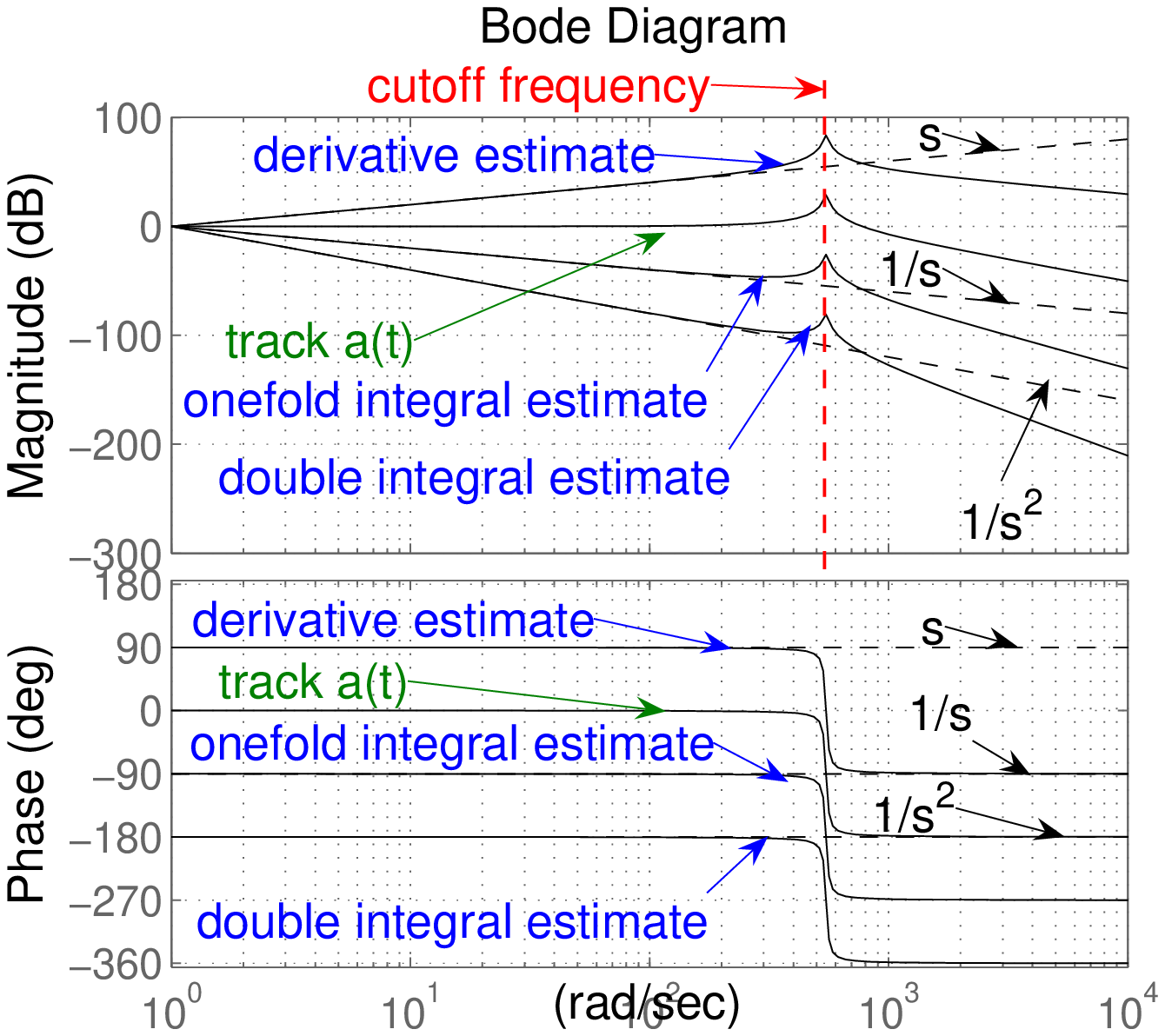}\\[0pt]
{\small 2(a) Bode plot when }$\varepsilon =0.1$\\[0pt]
\includegraphics[width=5.0in]{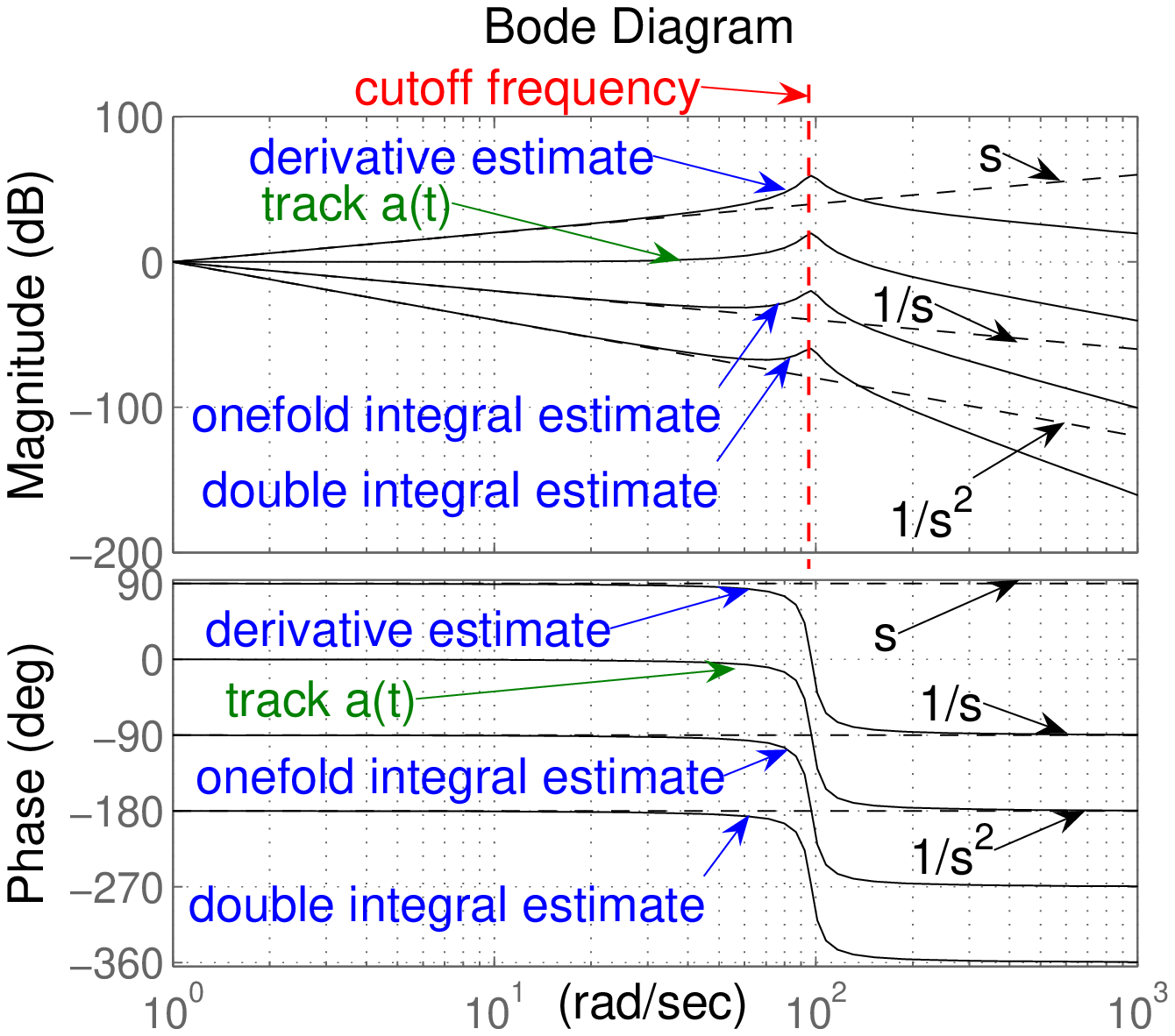}\\[0pt]
{\small 2(b) Bode plot when }$\varepsilon =0.2${\\[0pt]
{\small Figure 2 Bode plot of differentiation and double-integration observer%
}}
\end{center}
\end{figure}

\emph{3.1 Frequency characteristic with different perturbation parameter }$%
\varepsilon $

\emph{1) Differentiation-integration observer (22)}

For transfer function (14), let $n=3$, $p=2$, we obtain

\begin{equation}
\frac{X_{j}\left( s\right) }{A\left( s\right) }=\frac{k_{2}s^{j-1}}{%
\varepsilon ^{4}s^{3}+\varepsilon ^{3}k_{3}s^{2}+k_{2}s+\varepsilon k_{1}}%
,j\in \left\{ 1,2,3\right\}
\end{equation}%
where, $k_{1}=0.1$, $k_{2}=3$, $k_{3}=2$, Selecting $\varepsilon =0.1$ and $%
\varepsilon =0.2$, the Bode plots for the transfer function are described as
Figs. 1(a) and 1(b), respectively.

\emph{2) Differentiation and double-integration observer observer (26)}

For transfer function (14), let $n=4$, $p=3$, we obtain

\begin{equation}
\frac{X_{j}\left( s\right) }{A\left( s\right) }=\frac{k_{3}s^{j-1}}{%
\varepsilon ^{5}s^{4}+\varepsilon ^{4}k_{4}s^{2}+k_{3}s^{2}+\varepsilon
^{2}k_{2}s+\varepsilon k_{1}},j\in \left\{ 1,2,3,4\right\}
\end{equation}%
where, $k_{1}=0.01$, $k_{2}=0.1$, $k_{3}=3$, $k_{4}=2$, Selecting $%
\varepsilon =0.1$ and $\varepsilon =0.2$, the Bode plots for the transfer
function are described as Figs. 2(a) and 2(b), respectively.

Comparing with the ideal derivative operator $s$, the ideal\ integral
operator $1/s$ and $1/s^{2}$, not only the presented
differentiation-integration observers can obtain their estimations
accurately, but also the high-frequency noise is rejected sufficiently
(While in Figs. 1 and 2, the dash-lines represent the ideal operators and
the solid lines represent the proposed observers). From Figs. 1 and 2, after
the cutoff frequency lines, the estimations attenuate rapidly, and the
high-frequency noises are also reduced sufficiently. Parameter $\varepsilon $
affects the low-pass frequency bandwidth (See the cutoff frequency lines in
Figures 1 and 2): Decreasing the perturbation parameter $\varepsilon $, the
low-pass frequency bandwidth becomes larger, and the estimation speed
becomes fast; on the other hand, increasing perturbation parameter $%
\varepsilon $, the low-pass frequency bandwidth becomes smaller, and much
noise can be rejected sufficiently (See the cases of $\varepsilon =0.1$ and $%
\varepsilon =0.2$ in Figs.1 and 2, respectively).

\bigskip

\emph{3.2 The proposed rules of parameters selection}

For the differentiation-integration observers, there are some rules
suggested on the parameters selection:

1) The parameters $k_{i}$ $(i=1,\cdots ,n)$ decide the observer stability,
and they should be satisfied with the conditions in Lemma 1. Importantly,
the selection of $k_{i}$ $(i=1,\cdots ,n)$ should make the real parts of all
the eigenvalues of polynomial (5) negative for the small $\varepsilon \in
\left( 0,1\right) $.

\emph{a. For onefold integrator (20)}, the characteristic polynomial is $%
s^{2}+\frac{k_{2}/\varepsilon ^{2}}{\varepsilon }s+\frac{k_{1}}{\varepsilon
^{2}}$ (See Eq. (16) when $n=2$ and $p=2$). In fact, for $\varepsilon \in
\left( 0,1\right) $, the eigenvalues of the equivalent characteristic
polynomial $s^{2}+\frac{k_{2}}{\varepsilon ^{2}}s+k_{1}$ (See Eq. (17) when $%
n=2$ and $p=2$) can be written as the following form: $-a_{1}$, $-a_{2}$
(The real eigenvalues of the characteristic polynomial for observer (20) are
$-\frac{a_{1}}{\varepsilon }$, $-\frac{a_{2}}{\varepsilon }$). Therefore,
this polynomial can be written as

\begin{equation}
s^{2}+\frac{k_{2}}{\varepsilon ^{2}}%
s+k_{1}=(s+a_{1})(s+a_{2})=s^{2}+(a_{1}+a_{2})s+a_{1}a_{2}
\end{equation}%
By solving the above equation, it follows that

\begin{equation}
k_{1}=a_{1}a_{2},k_{2}=\varepsilon ^{2}(a_{1}+a_{2})  \notag
\end{equation}

From Eq. (14), the transfer function of the onefold integrator (20) can be
describled as

\begin{equation}
\frac{X_{2}\left( s\right) }{A\left( s\right) }=\frac{sk_{2}}{\varepsilon
^{3}s^{2}+sk_{2}+k_{1}\varepsilon }=\frac{sk_{2}/\varepsilon ^{3}}{%
s^{2}+sk_{2}/\varepsilon ^{3}+k_{1}/\varepsilon ^{2}}
\end{equation}%
Then its nature frequency is

\begin{equation}
\omega _{n}=\sqrt{k_{1}/\varepsilon ^{2}}
\end{equation}

Based on the requirement of filtering high-frequency noise, the perturbation
parameter can be selected as

\begin{equation}
\varepsilon =\sqrt{k_{1}/\omega _{n}^{2}}
\end{equation}

Because the drift is slow, the corresponding eigenvalue is selected to
approach the imaginary axis with respect to the other eigenvalue. For
example, let the eigenvalues be $-a_{1}=-100$, $-a_{2}=-0.02$, and select $%
\omega _{n}=8$. Then we obtain the observer parameters as follows:

\begin{eqnarray*}
k_{1} &=&100\times 0.02=2 \\
\varepsilon &=&0.1768 \\
k_{2} &=&(100+0.02)\times (2/64)=3.1256
\end{eqnarray*}

\bigskip

\emph{b. For the double integrator (24)} (when $n=3$ and $p=3$), the
characteristic polynomial is $s^{3}+\frac{k_{3}/\varepsilon ^{3}}{%
\varepsilon }s^{2}+\frac{k_{2}}{\varepsilon ^{2}}s+\frac{k_{1}}{\varepsilon
^{3}}$ (See Eq. (16) when $n=3$ and $p=3$). In fact, for $\varepsilon \in
\left( 0,1\right) $, the eigenvalues of this equivalent characteristic
polynomial $s^{3}+\frac{k_{3}}{\varepsilon ^{3}}s^{2}+k_{2}s+k_{1}$ (See Eq.
(17) when $n=3$ and $p=3$) can be written as the following form: $-a_{1}$, $%
-a_{21}+a_{22}i$, $-a_{21}-a_{22}i$ (The real eigenvalues of the
characteristic polynomial for observer (24) are $-\frac{a_{1}}{\varepsilon }$%
, $-\frac{a_{21}}{\varepsilon }+\frac{a_{22}}{\varepsilon }i$, $-\frac{a_{21}%
}{\varepsilon }-\frac{a_{22}}{\varepsilon }i$), where $a_{1},a_{21},a_{22}>0$%
. Two conjugate eigenvalues are supposed to exist in this polynomial.
Therefore, the polynomial $s^{3}+\frac{k_{3}}{\varepsilon ^{3}}%
s^{2}+k_{2}s+k_{1}$ can be written as

\begin{equation}
s^{3}+\frac{k_{3}}{\varepsilon ^{3}}%
s^{2}+k_{2}s+k_{1}=(s+a_{1})(s+a_{21}+a_{22}i)(s+a_{21}-a_{22}i)
\end{equation}%
By solving the above equation, it follows that

\begin{equation}
k_{1}=a_{1}(a_{21}^{2}+a_{22}^{2}),k_{2}=a_{21}^{2}+a_{22}^{2}+2a_{1}a_{21},k_{3}=\varepsilon ^{3}(a_{1}+2a_{21})
\end{equation}

It means that, after selecting the suitable eigenvalues $-a_{1}$, $%
-a_{21}+a_{22}i$, $-a_{21}-a_{22}i$ and $\varepsilon $ based on Bode plot
analysis, the parameters $k_{1}$, $k_{2}$ and $k_{3}$\ can be calculated.
Because the drifts are slow, the corresponding eigenvalues are selected to
approach the imaginary axis with respect to the other eigenvalues.

For example, selecting the eigenvalues of the polynomial as $-46.8218$, $%
-0.0266+0.0999i$, $-0.0266-0.0999i$, and $\varepsilon =0.4$,\ then $%
k_{1}=0.5 $, $k_{2}=2.5$, $k_{3}=3$; selecting the eigenvalues as $-15.6190$%
, $-0.0030+0.0800i$, $-0.0030-0.0800i$, and $\varepsilon =0.4$, then $%
k_{1}=0.1$, $k_{2}=0.1$, $k_{3}=1$. Obviously, the first parameters
selection has the stronger ability to correct the drift.

\bigskip

\emph{c. For differentiation-integration observer (22)} (when $n=3$ and $p=2$%
), the characteristic polynomial is $s^{3}+\frac{k_{3}}{\varepsilon }s^{2}+%
\frac{k_{2}/\varepsilon ^{2}}{\varepsilon ^{2}}s+\frac{k_{1}}{\varepsilon
^{3}}$ (See Eq. (16) when $n=3$ and $p=2$). In fact, for $\varepsilon \in
\left( 0,1\right) $, the eigenvalues of the equivalent characteristic
polynomial $s^{3}+k_{3}s^{2}+\frac{k_{2}}{\varepsilon ^{2}}s+k_{1}$ (See Eq.
(17) when $n=3$ and $p=2$) can be written as the following form: $%
-a_{11}+a_{12}i$, $-a_{11}-a_{12}i$, $-a_{2}$, (The real eigenvalues of the
characteristic polynomial for observer (22) are $-\frac{a_{11}}{\varepsilon }%
+\frac{a_{12}}{\varepsilon }i$, $-\frac{a_{11}}{\varepsilon }-\frac{a_{12}}{%
\varepsilon }i$, $-\frac{a_{2}}{\varepsilon }$), where $%
a_{11},a_{12},a_{1}>0 $. Two conjugate eigenvalues are supposed to exist in
this polynomial. Therefore, the polynomial $s^{3}+k_{3}s^{2}+\frac{k_{2}}{%
\varepsilon ^{2}}s+k_{1}$ can be written as

\begin{equation}
s^{3}+k_{3}s^{2}+\frac{k_{2}}{\varepsilon ^{2}}%
s+k_{1}=(s+a_{11}+a_{12}i)(s+a_{11}-a_{12}i)(s+a_{2})
\end{equation}%
By solving the above equation, it follows that

\begin{equation}
k_{1}=(a_{11}^{2}+a_{12}^{2})a_{2},k_{2}=\varepsilon
^{2}(a_{11}^{2}+a_{12}^{2}+2a_{11}a_{2}),k_{3}=2a_{11}+a_{2}
\end{equation}

It means that, after selecting the suitable eigenvalues $-a_{11}+a_{12}i$, $%
-a_{11}-a_{12}i$, $-a_{2}$ and $\varepsilon $ based on Bode plot analysis,
the parameters $k_{1}$, $k_{2}$ and $k_{3}$\ can be calculated.

\bigskip

2) In order to increase the estimation speed, $\varepsilon \in (0,1)$ should
decrease to make the low-pass frequency bandwidth larger; if much noise
exists, $\varepsilon $ should increase, the low-pass frequency bandwidth
becomes smaller, and the noise can be rejected sufficiently.

3) It is easy to see that the $k$-fold integrator provides for a much better
accuracy of $i$th-fold integral than the $l$-fold integrator, where, $k>l$
and $i=1,\cdots ,l-1$. For instance, the double integrator (24) provides for
a much better accuracy of onefold integral than the onefold integrator (20).

4) It is easy to see that the $k$th-order differentiator provides for a much
better accuracy of $i$th-order derivative than the $l$th-order
differentiator, where, $k>l$ and $i=1,\cdots ,l-1$. For instance, in the
high-order differentiator (18), the third-order differentiator (where $n=3$)
provides for a much better accuracy of first-order derivative than the
second-order differentiator (where $n=2$).

\section{Estimations by onefold integrator and double integrator}

In this section, we use the simulations to illustrate the effectiveness of
the proposed observers. The estimation performances of the presented
observers are compared with Extended Kalman Filter (EKF) [23], and a
long-time simulation is described to investigated their drift phenomena.

\emph{1) Estimation by onefold integrator (20)}

In this section, the onefold integrator (20), i.e.,

\begin{eqnarray*}
\dot{x}_{1} &=&x_{2} \\
\varepsilon ^{3}\dot{x}_{2} &=&-k_{1}\varepsilon x_{1}-k_{2}\left(
x_{2}-a\left( t\right) \right)
\end{eqnarray*}%
is used to estimate the integral from the signal $a(t)$ in spite of the
existence of stochastic non-zero mean noise $\delta (t)$ and measurement
error $d(t)$.

\begin{figure}[H]
\begin{center}
\includegraphics[width=2.90in]{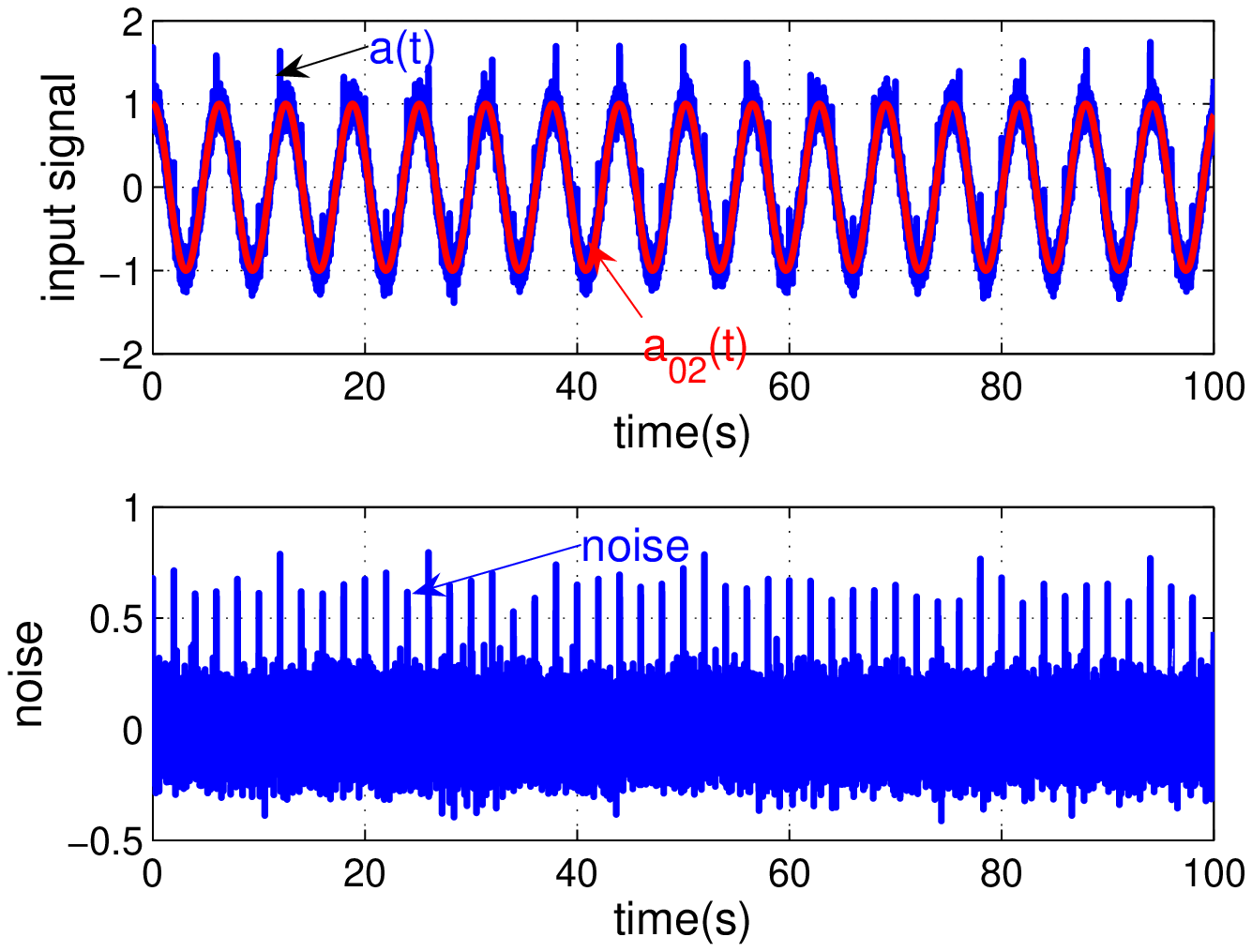}\\[0pt]
{\small 3(a) Input signal}\\[0pt]
\includegraphics[width=2.90in]{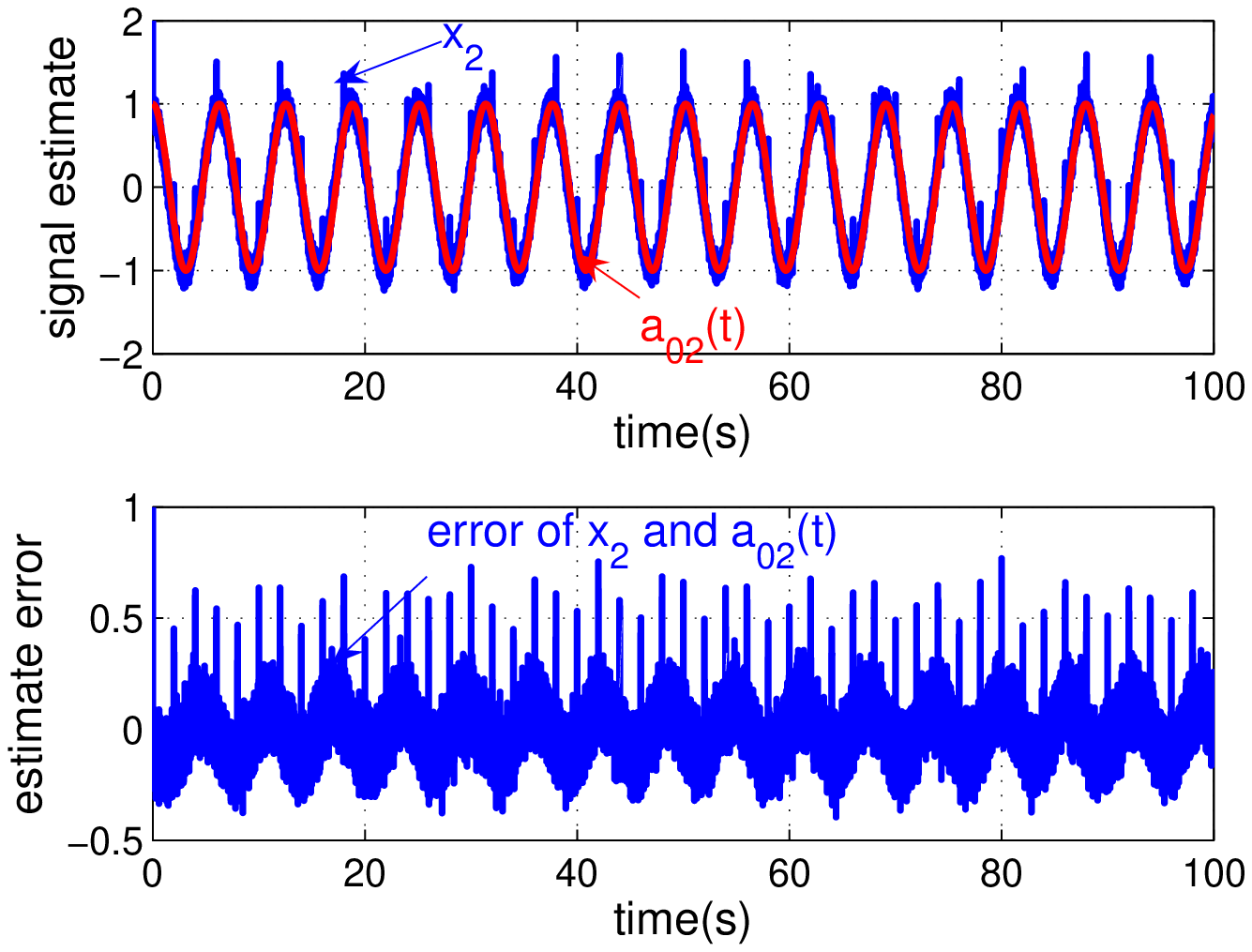}\\[0pt]
{\small 3(b) Signal estimate\\[0pt]
\includegraphics[width=2.90in]{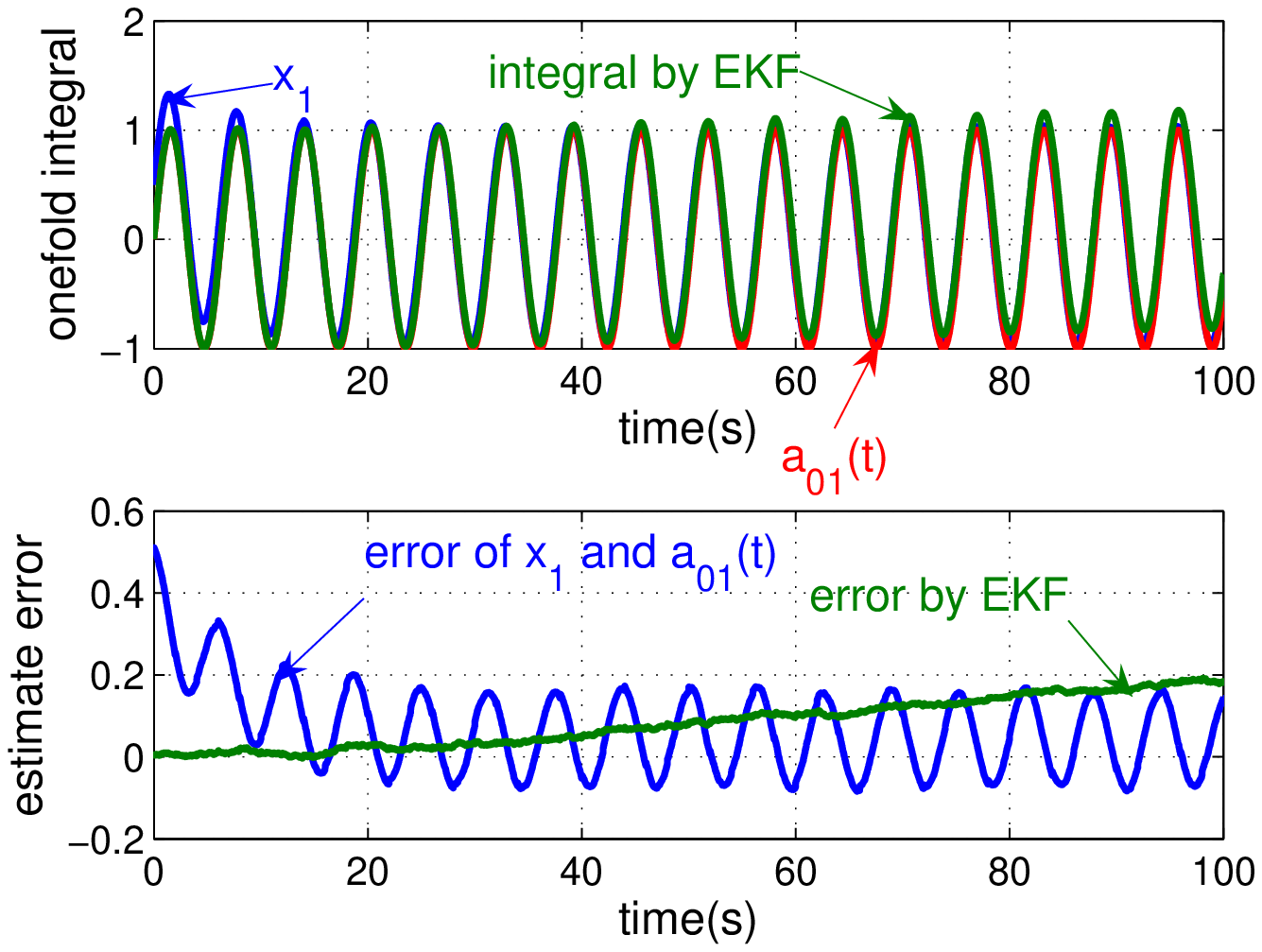}\\[0pt]
3(c) Onefold integral estimate}\\[0pt]
{\small Figure 3 Estimation by onefold integrator (19) in 100s}\\[0pt]
\includegraphics[width=2.90in]{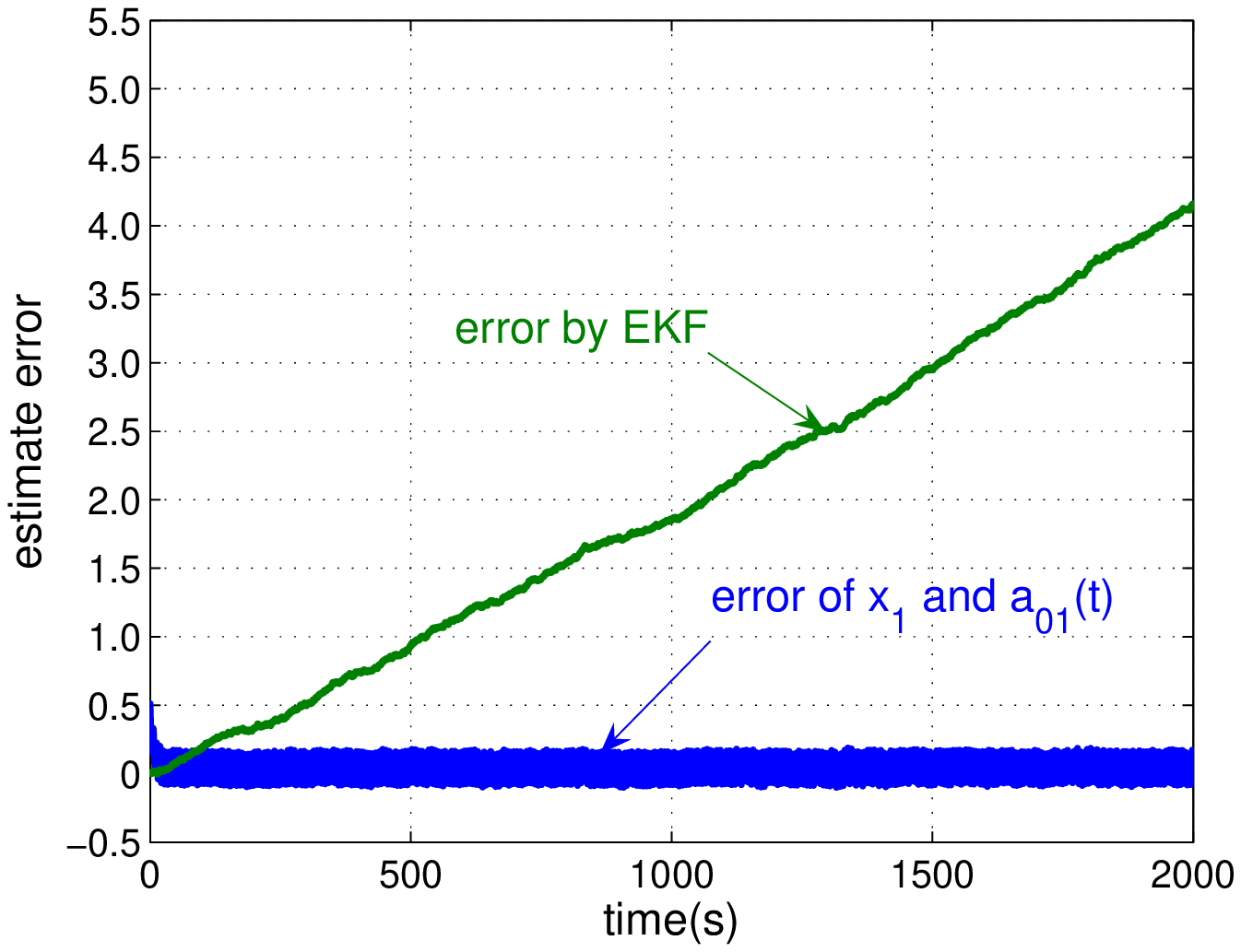}\\[0pt]
{\small Figure 4 Comparison of onefold integrator (19) and EKF in 2000s}
\end{center}
\end{figure}

\begin{figure}[H]
\begin{center}
\includegraphics[width=2.90in]{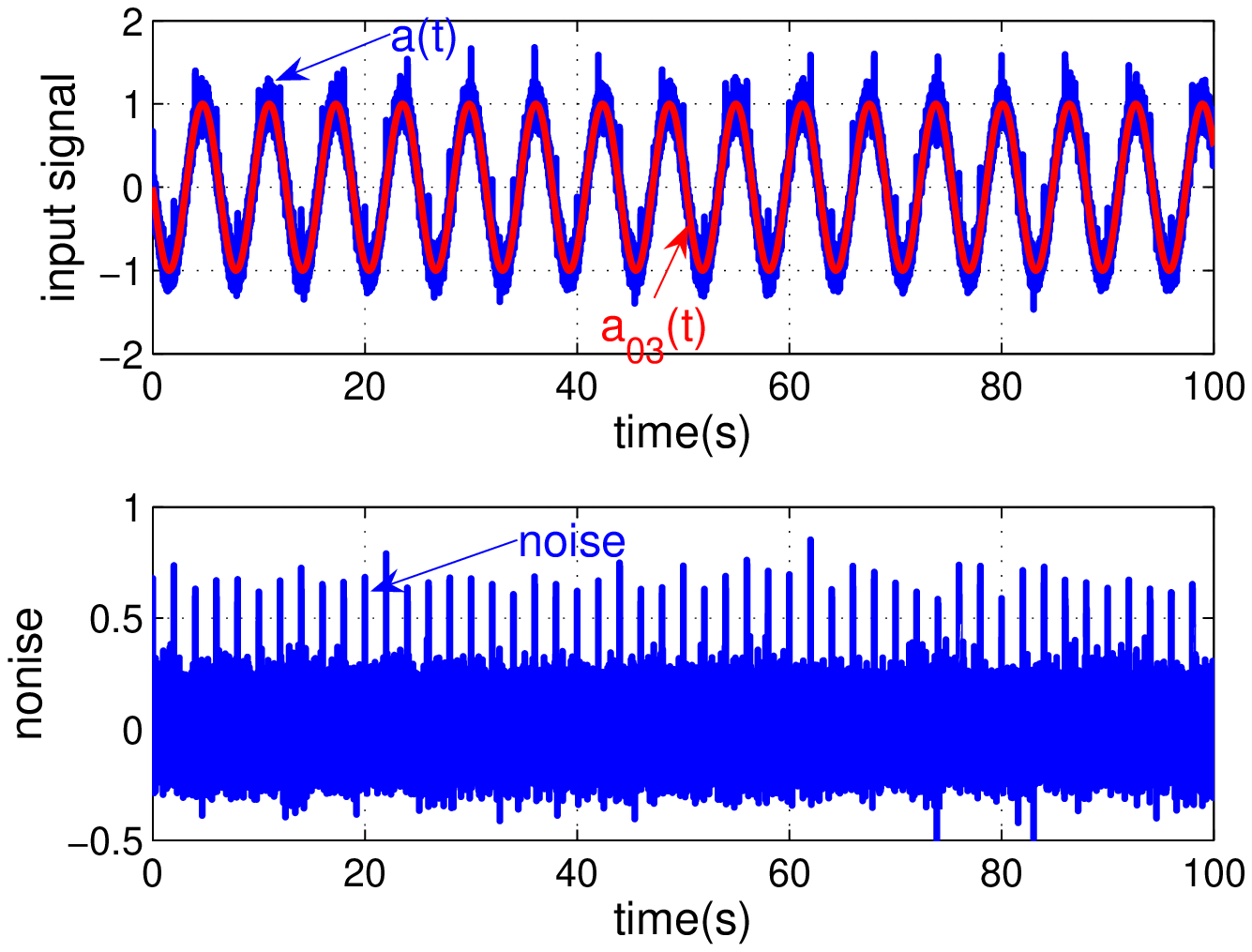}\\[0pt]
{\small 5(a) Input signal}\\[0pt]
\includegraphics[width=2.90in]{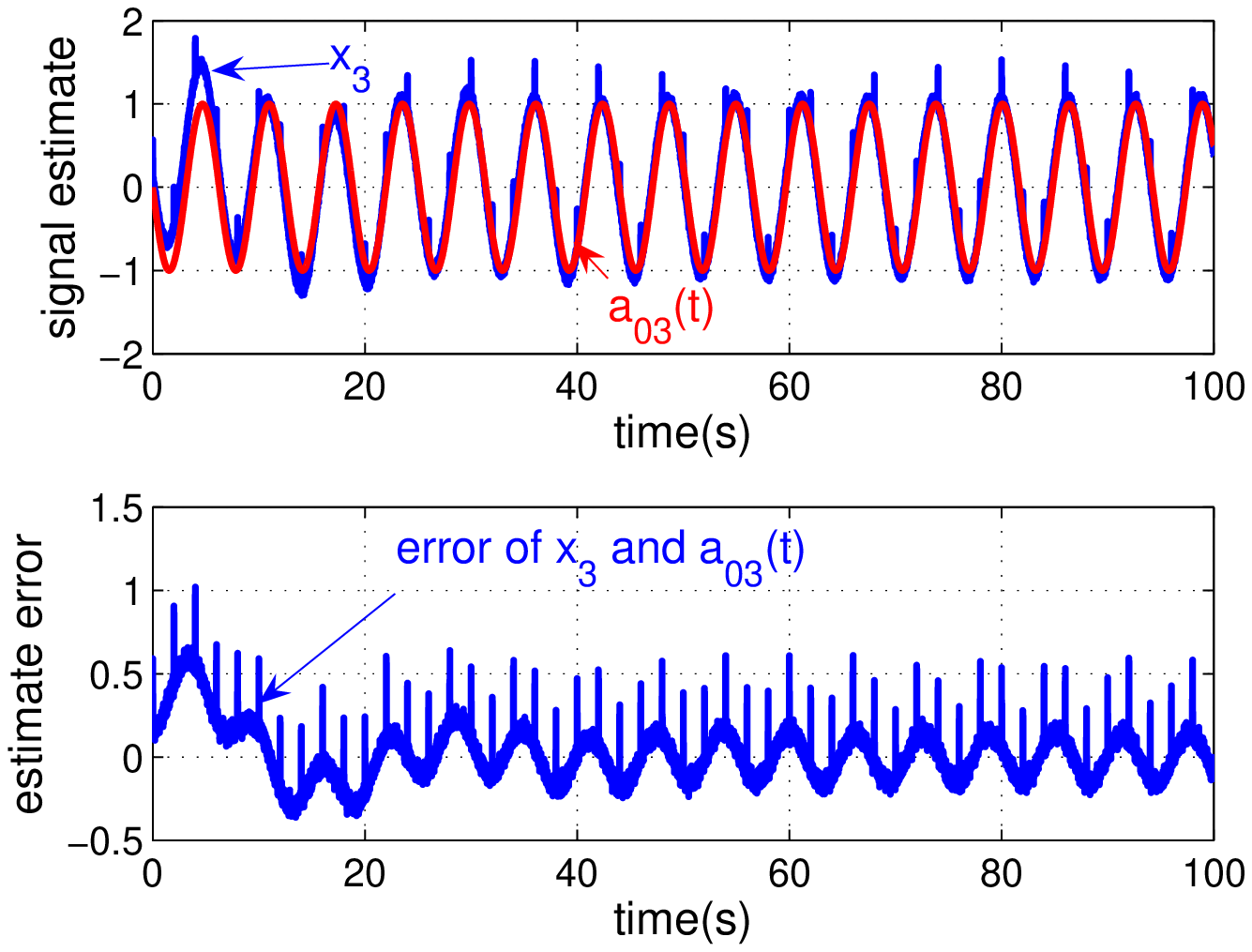}\\[0pt]
{\small 5(b) Signal estimate\\[0pt]
\includegraphics[width=2.90in]{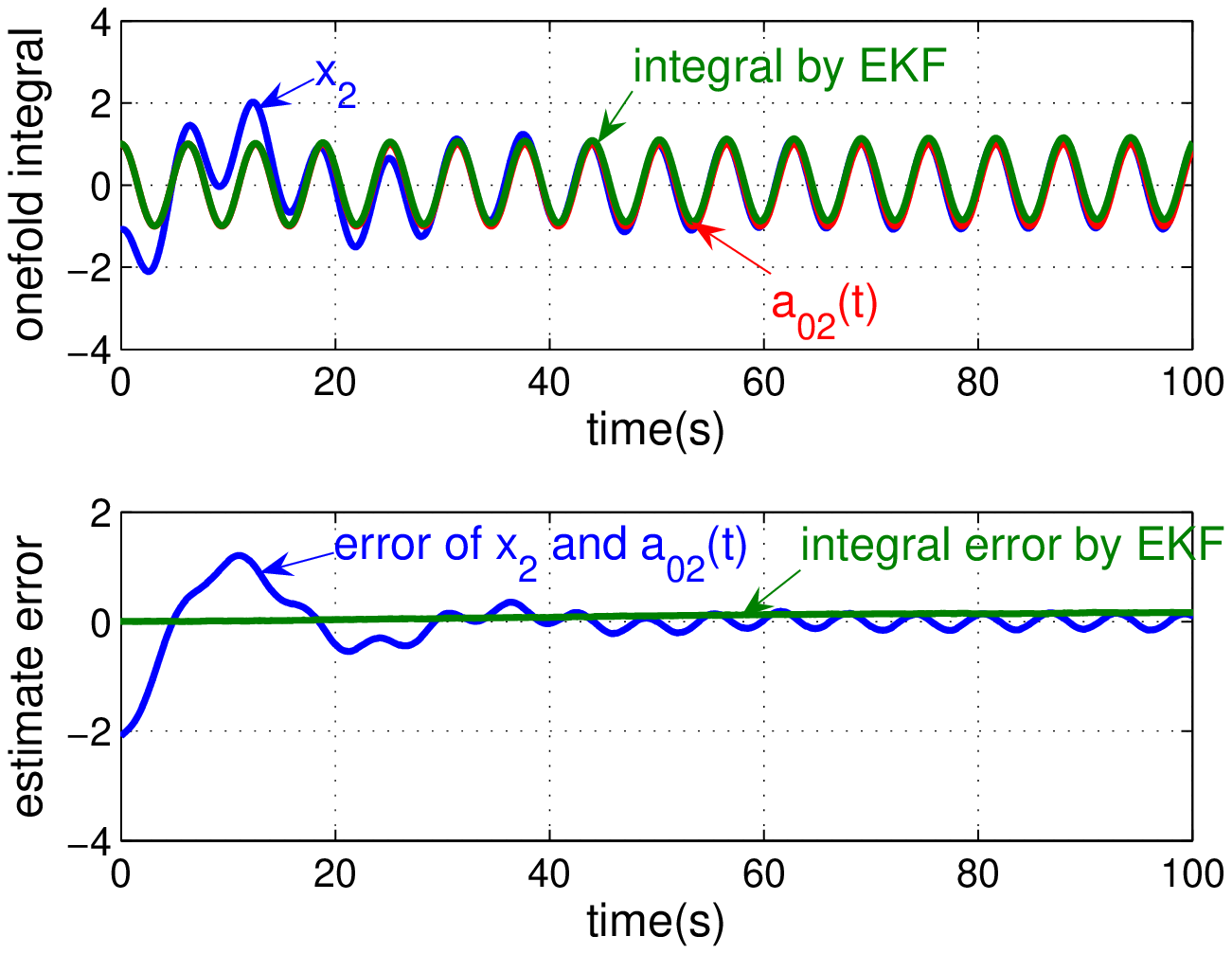}\\[0pt]
5(c) Onefold integral estimate\\[0pt]
\includegraphics[width=2.90in]{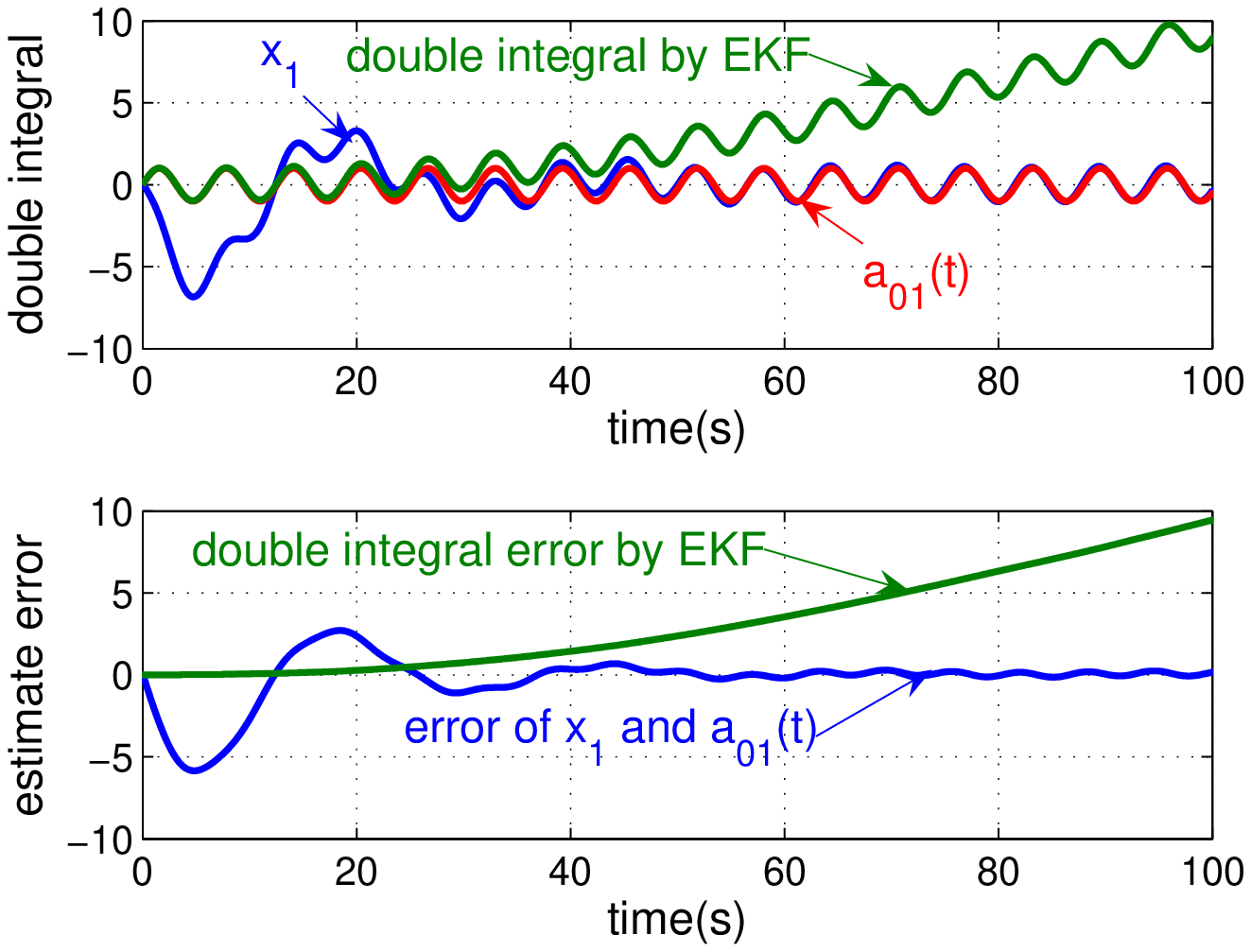}\\[0pt]
5(d) Double integral estimate}\\[0pt]
{\small Figure 5 Estimation by double integrator (23) in 100s}
\end{center}
\end{figure}

\begin{figure}[H]
\begin{center}
\includegraphics[width=2.90in]{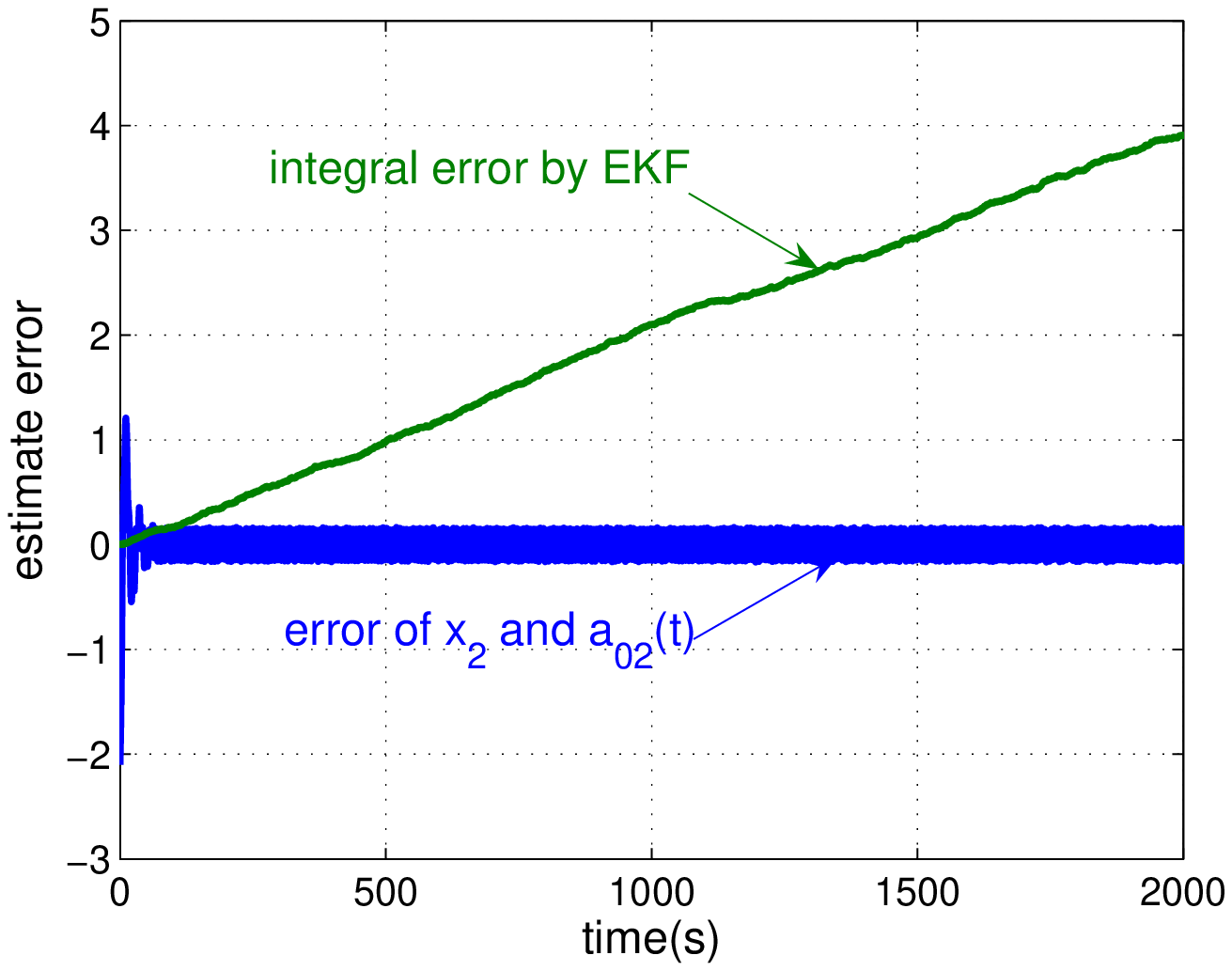}\\[0pt]
{\small 6(a) Onefold integral estimate}\\[0pt]
\includegraphics[width=2.90in]{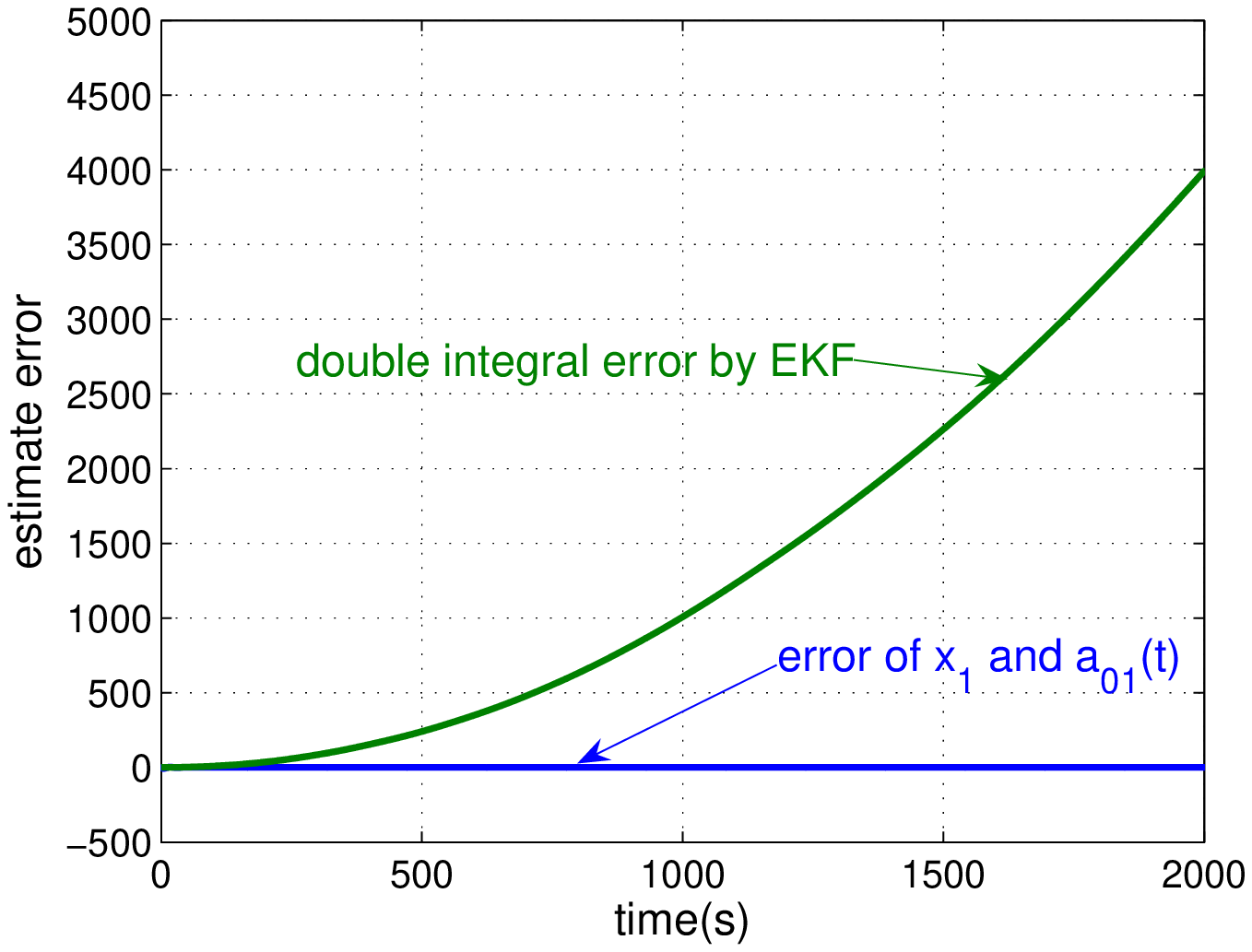}\\[0pt]
{\small 6(a) Double integral estimate}\\[0pt]
{\small 6 Comparison of double integrator (23) and EKF in 2000s}
\end{center}
\end{figure}

Here, the stochastic non-zero mean noise is selected, and the mean value of
the noise is not equal to zero (See the noise in Fig. 3(a)). The non-zero
mean noise $\delta (t)$ consists of following two signals: Random number
with Mean=0, Variance=0.01, Initial speed=0, and Sample time=0; Pulses with
Amplitude=0.5, Period=2s, Pulse width=1, and Phase delay=0.

The signal $a_{02}(t)=\cos (t)$ is selected as the reference signal, and $%
a\left( t\right) =a_{02}(t)+\delta (t)+d(t)$. Therefore, $%
a_{01}=\int_{0}^{t}a\left( \tau \right) d\tau =\sin (t)$. Integrator
parameters: $k_{1}=2$, $k_{2}=2.7783$, $\varepsilon =0.1667$. Suppose the
initial state is $(x_{1}(0),x_{2}(0))=(0.5,2)$. In the onefold integrator
(20), $x_{2}$ estimates signal $a_{02}(t)$, $x_{1}$ estimate the onefold
integral $a_{01}(t)$. Signal $a_{02}(t)$ tracking, the onefold integral
estimation in 100 seconds are presented in Fig. 3. Fig. 3(a) provides signal
$a_{02}\left( t\right) $ with stochastic noise. Fig. 3(b) describes signal $%
a_{02}(t)$ estimation. Fig. 3(c) presents the comparison of onefold integral
estimation by onefold integrator (20) and Extended Kalman filter [23]. Figs.
4 describes the estimation comparison in 2000 seconds.

From Figs. 3(c) and 4, the obvious estimation drift of onefold integral
exists by the Extended Kalman filter. With respect to the Extended Kalman
filter, the proposed onefold integrator (20) showed the promising estimation
ability and robustness in spite of the existence of the non-zero mean
stochastic noise. Furthermore, from Fig. 4, no drift phenomenon happened in
the long-time estimation.\newpage

\emph{2) Estimation by double integrator (24)}

In this section, the double integrator (24), i.e.,

\begin{eqnarray*}
\dot{x}_{1} &=&x_{2} \\
\dot{x}_{2} &=&x_{3} \\
\varepsilon ^{4}\dot{x}_{3} &=&-k_{1}\varepsilon x_{1}-k_{2}\varepsilon
^{2}x_{2}-k_{3}\left( x_{3}-a\left( t\right) \right)
\end{eqnarray*}%
is used to estimate the onefold and double integrals from the signal $a(t)$
in spite of the existence of stochastic non-zero mean noise $\delta (t)$ and
measurement error $d(t)$.

Here, the stochastic non-zero mean noise in 1) is selected.

The signal $a_{03}(t)=-\sin (t)$ is selected as the reference signal, and $%
a\left( t\right) =a_{03}(t)+\delta (t)+d(t)$. Therefore, $%
a_{02}=\int_{0}^{t}a_{03}(\sigma )d\sigma =\cos (t)$, and $%
a_{01}=\int_{0}^{t}\int_{0}^{s}a\left( \sigma \right) d\sigma d\tau =\sin
(t) $.

The double integrator parameters: $k_{1}=0.5$, $k_{2}=2.5$, $k_{3}=3$, $%
\varepsilon =0.4$. Suppose the initial state is $%
(x_{1}(0),x_{2}(0),x_{3}(0))=(0.1,-1.1,0.1)$. In the double integrator (24),
$x_{3}$ tracks signal $a_{03}(t)$, $x_{2}$ and $x_{1}$ estimate the onefold
and double integrals of signal $a_{03}(t)$, respectively.

Signal $a_{03}(t)$ tracking, the onefold and double integral estimations in
100 seconds are presented in Fig. 5. Fig. 5(a) provides signal $a_{03}\left(
t\right) $ with stochastic noise. Fig. 5(b) describes $a_{03}(t)$
estimation. Figs. 5(c) and 5(d) present the comparisons of onefold and
double integral estimations by the double integrator (24) and the Extended
Kalman filter [23]. Figs. 6(a)-6(b) describe the estimation comparisions of
in 2000 seconds.

From Figs. 5(c), 5(d), 6(a) and 6(b), the obvious estimation drifts of
onefold and double integrals exist by the Extended Kalman filter. With
respect to the Extended Kalman filter, despite the existence of the
intensive non-zero mean stochastic noise, the proposed double integrator
(24) showed the promising estimation ability and robustness. Furthermore,
from Figs. 6(a)-6(b), no drift phenomenon happened in the long-time
estimations.

\section{Application to quadrotor aircraft}

In this paper, the mathematical model and reference trajectory of the
quadrotor aircraft described in [29] are used. The description of forces and
torques of the quadrotor aircraft is shown in Fig. 7 [29].

\begin{figure}[H]
\begin{center}
\includegraphics[width=2.00in]{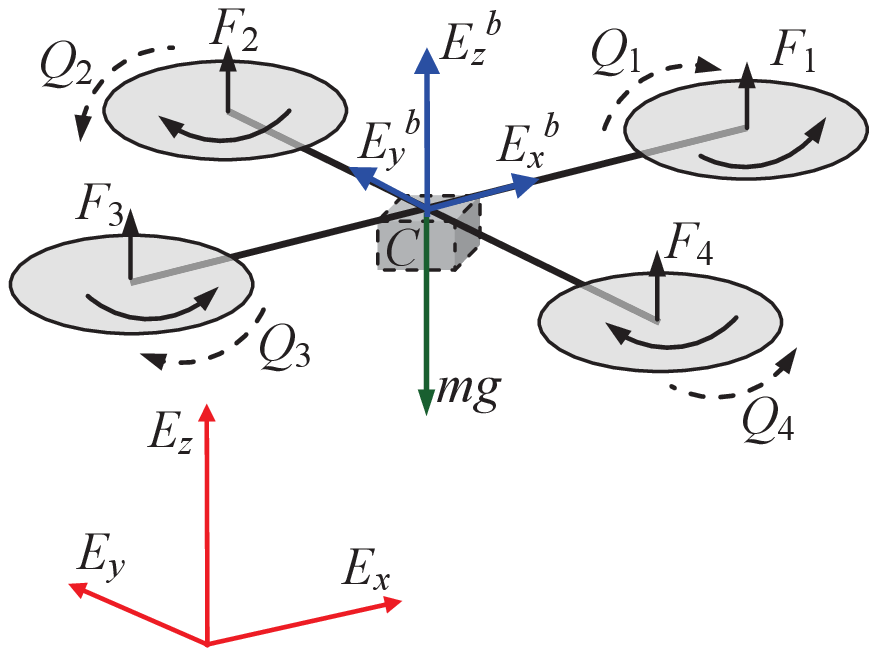}\\[0pt]
{Figure 7 Forces and torques of quadrotor aircraft}\\[0pt]
\end{center}
\end{figure}

Let $\Xi _{g}=\left( E_{x},E_{y},E_{z}\right) $ denote the right handed
inertial frame, and $\Xi _{b}=\left( E_{x}^{b},E_{y}^{b},E_{z}^{b}\right) $
denote the frame attached to the aircraft's fuselage whose origin is at the
center of gravity. ($\psi ,\theta ,\phi $) denotes the aircraft orientation
expressed in the yaw, pitch and roll angles (Euler angles). The symbol $%
c_{\theta }$ is used for $\cos \theta $ and $s_{\theta }$ for $\sin \theta $%
. $R_{bg}$ is the transformation matrix from the frame $\Xi _{b}$ to $\Xi
_{g}$, and

\begin{equation}
R_{bg}=\left[
\begin{array}{ccccc}
c_{\psi }c_{\theta } &
\begin{array}{cc}
&
\end{array}
& s_{\psi }c_{\phi }+c_{\psi }s_{\theta }s_{\phi } &
\begin{array}{cc}
&
\end{array}
& s_{\psi }s_{\phi }-c_{\psi }s_{\theta }c_{\phi } \\
-s_{\psi }c_{\theta } &
\begin{array}{cc}
&
\end{array}
& c_{\psi }c_{\phi }-s_{\psi }s_{\theta }s_{\phi } &
\begin{array}{cc}
&
\end{array}
& c_{\psi }s_{\phi }+s_{\psi }s_{\theta }c_{\phi } \\
s_{\theta } &
\begin{array}{cc}
&
\end{array}
& -c_{\theta }s_{\phi } &
\begin{array}{cc}
&
\end{array}
& c_{\theta }c_{\phi }%
\end{array}%
\right]
\end{equation}

For the quadrotor aircraft, the right--left rotors rotate clockwise and the
front-rear ones rotate counterclockwise (See Fig. 7). The rotational
directions of the rotors do not change (i.e., $\omega _{i}>0$, $i\in
\{1,2,3,4\}$). The reactive torque generated by the rotor $i$ due to the
rotor drag is $Q_{i}=k\omega _{i}^{2}$, and the total thrust generated by
the four rotors is $F=\sum\limits_{{i=1}}^{{4}}F_{i}=b\sum\limits_{{i=1}}^{{4%
}}\omega _{i}^{2}$, where $F_{i}=b\omega _{i}^{2}$ is the lift generated by
the rotor $i$ in free air, and $k,b>0$ are two parameters depending on the
density of air, the size, shape, and pitch angle of the blades, as well as
other factors. Therefore, we obtain $Q_{i}=\frac{k}{b}F_{i},i=1,2,3,4$. Thus
the sum reactive torque generated by the four rotors due to the rotor drags
is $Q=$ $\sum\limits_{{i=1}}^{{4}}(-1)^{i}Q_{i}=\frac{k}{b}\sum\limits_{{i=1}%
}^{{4}}(-1)^{i}F_{i}$.

The motion equations in the coordinate ($x,y,z$) are then [29]

\begin{eqnarray}
m\ddot{x} &=&(s_{\psi }s_{\phi }-c_{\psi }s_{\theta }c_{\phi })F-k_{x}\dot{x}%
+\delta _{x}  \notag \\
m\ddot{y} &=&(c_{\psi }s_{\phi }+s_{\psi }s_{\theta }c_{\phi })F-k_{y}\dot{y}%
+\delta _{y}  \notag \\
m\ddot{z} &=&c_{\theta }c_{\phi }F-mg-k_{z}\dot{z}+\delta _{z} \\
J_{z}\ddot{\psi} &=&\frac{k}{b}\sum\limits_{{i=1}}^{{4}}(-1)^{i}F_{i}-k_{%
\psi }\dot{\psi}+\delta _{\psi }  \notag \\
J_{y}\ddot{\theta} &=&(F_{1}-F_{3})l-lk_{\theta }\dot{\theta}+\delta
_{\theta }  \notag \\
J_{x}\ddot{\phi} &=&(F_{2}-F_{4})l-lk_{\phi }\dot{\phi}+\delta _{\phi }
\end{eqnarray}%
where, $m$ is the mass of the aircraft; $g$ is the gravity acceleration; $%
J_{x}$, $J_{y}$ and $J_{z}$ are the three-axis moment of inertias; $k_{x}$, $%
k_{y}$, $k_{z}$, $k_{\psi }$, $k_{\theta }$ and $k_{\phi }$ are the drag
coefficients; $l$ is the distance between each rotor and the center of
gravity. $\delta _{x}$, $\delta _{y}$ and $\delta _{z}$ are the bounded
disturbances and uncertainties in position dynamics; $\delta _{\psi }$, $%
\delta _{\theta }$ and $\delta _{\phi }$ are the bounded disturbances and
uncertainties in attitude dynamics.

Here, for the quadrotor aircraft, we are interested in designing the
observers to estimate ($\dot{x}$, $\dot{y}$, $\dot{z}$, $\psi $, $\theta $, $%
\phi )$ and the uncertainties ($k_{x}$, $k_{y}$, $k_{z}$, $k_{\psi }$, $%
k_{\theta }$, $k_{\phi }$) and ($\delta _{x}$, $\delta _{y}$, $\delta _{z}$,
$\delta _{\psi }$, $\delta _{\theta }$, $\delta _{\phi }$) from the
information of ($x$, $y$, $z$, $\dot{\psi}$, $\dot{\theta}$, $\dot{\phi}$).
Moreover, based on these observers, the controllers $F_{i}$ ($i=1,2,3,4)$
will be designed to implement: $x\rightarrow x_{d}$, $\dot{x}\rightarrow
\dot{x}_{d}$, $y\rightarrow y_{d}$, $\dot{y}\rightarrow \dot{y}_{d}$, $%
z\rightarrow z_{d}$, $\dot{z}\rightarrow \dot{z}_{d}$, and $\psi \rightarrow
\psi _{d}$, $\dot{\psi}\rightarrow \dot{\psi}_{d}$, $\theta \rightarrow
\theta _{d}$, $\dot{\theta}\rightarrow \dot{\theta}_{d}$, $\phi \rightarrow
\phi _{d}$, $\phi \rightarrow \dot{\phi}_{d}$ as $t\rightarrow \infty $.

\bigskip

\emph{4.1 Observer designs for the quadrotor aircraft}

For quadrotor aircraft equations (39) and (40), we consider that ($\dot{x}$,
$\dot{y}$, $\dot{z}$, $\psi $, $\theta $, $\phi $) is not measured directly,
($k_{x}$, $k_{y}$, $k_{z}$, $k_{\psi }$, $k_{\theta }$, $k_{\phi }$) and ($%
\delta _{x}$, $\delta _{y}$, $\delta _{z}$, $\delta _{\psi }$, $\delta
_{\theta }$, $\delta _{\phi }$) are bounded and unknown. Select the
auxiliary controller vector as

\begin{equation}
u_{p}=\left[
\begin{array}{c}
u_{px} \\
u_{py} \\
u_{pz}%
\end{array}%
\right] =\left[
\begin{array}{c}
s_{\psi }s_{\phi }-c_{\psi }s_{\theta }c_{\phi } \\
c_{\psi }s_{\phi }+s_{\psi }s_{\theta }c_{\phi } \\
c_{\theta }c_{\phi }%
\end{array}%
\right] F
\end{equation}%
Then we can find that

\begin{equation}
F=\left\Vert u_{p}\right\Vert _{2}=\sqrt{u_{px}^{2}+u_{py}^{2}+u_{pz}^{2}}
\end{equation}%
That is to say, after designing ($u_{px},u_{py},u_{pz}$), $F$ can be
calculated. Therefore, ($u_{px},u_{py},u_{pz}$) is known. Let [29]

\begin{eqnarray}
h_{1}(t) &=&\frac{u_{px}}{m},h_{2}(t)=\frac{u_{py}}{m},h_{3}(t)=\frac{u_{pz}%
}{m}-g,  \notag \\
h_{4}(t) &=&\frac{k}{J_{z}b}\sum\limits_{{i=1}}^{{4}}(-1)^{i}F_{i},h_{5}(t)=%
\frac{l}{J_{y}}(F_{1}-F_{3}),h_{6}(t)=\frac{l}{J_{x}}(F_{2}-F_{4})
\end{eqnarray}

\begin{eqnarray}
d_{1}(t) &=&(\delta _{x}-k_{x}\dot{x})/m,d_{2}(t)=(\delta _{y}-k_{y}\dot{y}%
)/m,  \notag \\
d_{3}(t) &=&(\delta _{z}-k_{z}\dot{z})/m,d_{4}(t)=(\delta _{\psi }-k_{\psi }%
\dot{\psi})/J_{z},  \notag \\
d_{5}(t) &=&(\delta _{\theta }-lk_{\theta }\dot{\theta})/J_{y},d_{6}(t)=(%
\delta _{\phi }-lk_{\phi }\dot{\phi})/J_{x}
\end{eqnarray}

\begin{eqnarray}
w_{1,1} &=&x,w_{2,1}=y,w_{3,1}=z,w_{4,1}=\psi ,w_{5,1}=\theta ,w_{6,1}=\phi
\notag \\
w_{1,2} &=&\dot{x},w_{2,2}=\dot{y},w_{3,2}=\dot{z},w_{4,2}=\dot{\psi}%
,w_{5,2}=\dot{\theta},w_{6,2}=\dot{\phi}
\end{eqnarray}%
then the position dynamics (39) can be rewritten as

\begin{eqnarray}
\dot{w}_{i,1} &=&w_{i,2}  \notag \\
\dot{w}_{i,2} &=&h_{i}(t)+d_{i}(t)  \notag \\
y_{opi} &=&w_{i,1}
\end{eqnarray}%
where $i=1,2,3$, and the attitude dynamics (40) can be given by

\begin{eqnarray}
\dot{w}_{i,1} &=&w_{i,2}  \notag \\
\dot{w}_{i,2} &=&h_{i}(t)+d_{i}(t)  \notag \\
y_{opi} &=&w_{i,2}
\end{eqnarray}%
where $i=4,5,6$.

Based on Theorem 1 and Lemma 1, the following corollaries describe the the
observer designs for the quadrotor aircraft.\newpage

\emph{1) The third-order differentiator (18) for velocity and uncertainties
estimate in position dynamics}

The following corollary gives the observers to estimate ($\dot{x},\dot{y},%
\dot{z}$) and uncertainties in the position dynamics.

\emph{Corollary 2:} The observer (18) (where $n=3$) are designed for
aircraft position dynamics (39) as follows:

\begin{eqnarray}
\dot{x}_{i,1} &=&x_{i,2}  \notag \\
\dot{x}_{i,2} &=&x_{i,3}  \notag \\
\varepsilon _{i}^{3}\dot{x}_{i,3} &=&-k_{1}\left( x_{i,1}-w_{i,1}\right)
-k_{2}\varepsilon _{i}x_{i,2}-k_{3}\varepsilon _{i}^{2}x_{i,3}
\end{eqnarray}%
where $i=1,2,3$. From $(x,y,z)$, we can estimate $(\dot{x},\dot{y},\dot{z})$
and $d_{i}(t)$ ($i=1,2,3$) by the differentiators in Eq. (48), and the
following conclusions hold:

\begin{eqnarray}
\underset{\varepsilon \rightarrow 0}{\lim }x_{1,1} &=&x,\underset{%
\varepsilon \rightarrow 0}{\lim }x_{1,2}=\dot{x},\underset{\varepsilon
\rightarrow 0}{\lim }x_{1,3}-h_{1}(t)=d_{1}(t)  \notag \\
\underset{\varepsilon \rightarrow 0}{\lim }x_{2,1} &=&y,\underset{%
\varepsilon \rightarrow 0}{\lim }x_{2,2}=\dot{y},\underset{\varepsilon
\rightarrow 0}{\lim }x_{2,3}-h_{2}(t)=d_{2}(t)  \notag \\
\underset{\varepsilon \rightarrow 0}{\lim }x_{3,1} &=&z,\underset{%
\varepsilon \rightarrow 0}{\lim }x_{3,2}=\dot{z},\underset{\varepsilon
\rightarrow 0}{\lim }x_{3,3}-h_{3}(t)=d_{3}(t)
\end{eqnarray}

\bigskip

\emph{2) Differentiation-integration observer (22) for attitude angle
estimation}

The following corollary gives the observers to estimate ($\psi ,\theta ,\phi
$) and uncertainties in the attitude dynamics.

\emph{Corollary 3:} The differentiation-integrator observers are designed
for aircraft attitude dynamics (40) as follows:

\begin{eqnarray}
\dot{x}_{i,1} &=&x_{i,2}  \notag \\
\dot{x}_{i,2} &=&x_{i,3}  \notag \\
\varepsilon _{i}^{4}\dot{x}_{i,3} &=&-k_{i,1}\varepsilon
_{i}x_{i,1}-k_{i,2}\left( x_{i,2}-w_{i,2}\right) -k_{i,3}\varepsilon
_{i}^{3}x_{i,3}
\end{eqnarray}%
where $i=4,5,6$. From $(\dot{\psi},\dot{\theta},\dot{\phi})$, we can
estimate $(\psi ,\theta ,\phi )$ and $d_{i}(t)$ ($i=4,5,6$) by the
differentiation-integration observers in Eq. (50), and the following
conclusions hold:

\begin{eqnarray}
\underset{\varepsilon \rightarrow 0}{\lim }x_{4,1} &=&\psi ,\underset{%
\varepsilon \rightarrow 0}{\lim }x_{4,2}=\dot{\psi},\underset{\varepsilon
\rightarrow 0}{\lim }x_{4,3}-h_{4}(t)=d_{4}(t)  \notag \\
\underset{\varepsilon \rightarrow 0}{\lim }x_{5,1} &=&\theta ,\underset{%
\varepsilon \rightarrow 0}{\lim }x_{5,2}=\dot{\theta},\underset{\varepsilon
\rightarrow 0}{\lim }x_{5,3}-h_{5}(t)=d_{5}(t)  \notag \\
\underset{\varepsilon \rightarrow 0}{\lim }x_{6,1} &=&\phi ,\underset{%
\varepsilon \rightarrow 0}{\lim }x_{6,2}=\dot{\phi},\underset{\varepsilon
\rightarrow 0}{\lim }x_{6,3}-h_{6}(t)=d_{6}(t)
\end{eqnarray}

\emph{4.2 Controller design}

In this section, a control law will be designed for the attitude
stabilization and trajectory tracking. The unknown states and uncertainties
are reconstructed by the presented observers. Suppose the reference
trajectory and its finite-order derivatives are bounded, and they can be
generated directly.

For reference trajectory ($x_{d},y_{d},z_{d}$), define $e_{1}=x-x_{d}$, $%
e_{2}=\dot{x}-\dot{x}_{d}$, $e_{3}=y-y_{d}$, $e_{4}=\dot{y}-\dot{y}_{d}$, $%
e_{5}=z-z_{d}$, and $e_{6}=\dot{z}-\dot{z}_{d}$. The system error for
position dynamics (39) can be written as

\begin{equation}
\ddot{e}_{p}=m^{-1}(u_{p}+\Xi _{p}+\delta _{p})
\end{equation}%
where

\begin{equation}
{\small e}_{p}{\small =}\left[
\begin{array}{ccc}
{\small e}_{1} & {\small e}_{3} & {\small e}_{5}%
\end{array}%
\right] ^{T},{\small \Xi _{p}}{\small =}\left[
\begin{array}{c}
-m\ddot{x}_{d} \\
-m\ddot{y}_{d} \\
-m\ddot{z}_{d}-mg%
\end{array}%
\right] ,{\small \delta _{p}=}\left[
\begin{array}{c}
{\small \delta }_{x}{\small -k}_{x}{\small \dot{x}} \\
{\small \delta }_{y}{\small -k}_{y}{\small \dot{y}} \\
{\small \delta }_{z}{\small -k}_{z}{\small \dot{z}}%
\end{array}%
\right]
\end{equation}

For reference attitude angle ($\psi _{d},\theta _{d},\phi _{d}$), define $%
e_{7}=\psi -\psi _{d}$, $e_{8}=\dot{\psi}-\dot{\psi}_{d}$, $e_{9}=\theta
-\theta _{d}$, $e_{10}=\dot{\theta}-\dot{\theta}_{d}$, $e_{11}=\phi -\phi
_{d}$, $e_{12}=\dot{\phi}-\dot{\phi}_{d}$. The system error for attitude
dynamics (40) is given by

\begin{equation}
\ddot{e}_{a}=J^{-1}(u_{a}+\Xi _{a}+\delta _{a})
\end{equation}%
where

\begin{equation}
e_{a}=\left[
\begin{array}{c}
e_{7} \\
e_{9} \\
e_{11}%
\end{array}%
\right] ,u_{a}=\left[
\begin{array}{c}
\frac{k}{b}\sum\limits_{{i=1}}^{{4}}(-1)^{i}F_{i} \\
(F_{1}-F_{3})l \\
(F_{2}-F_{4})l%
\end{array}%
\right] ,\Xi _{a}=\left[
\begin{array}{c}
-J_{z}\ddot{\psi}_{d} \\
-J_{y}\ddot{\theta}_{d} \\
-J_{x}\ddot{\phi}_{d}%
\end{array}%
\right] ,\delta _{a}=\left[
\begin{array}{c}
\delta _{\psi }-k_{\psi }\dot{\psi} \\
\delta _{\theta }-lk_{\theta }\dot{\theta} \\
\delta _{\phi }-lk_{\phi }\dot{\phi}%
\end{array}%
\right] ,J=diag\{J_{z},J_{y},J_{x}\}
\end{equation}

\bigskip

\emph{4.2.1 Controller design for position dynamics}

\emph{Theorem 2:} For the position dynamics (39), to track the reference
trajectory ($x_{d},y_{d},z_{d}$), if the controller is selected as

\begin{equation}
u_{p}=-\Xi _{p}-\widehat{\delta }_{p}-m(k_{p1}\widehat{e}_{p}+k_{p2}\widehat{%
\dot{e}}_{p})
\end{equation}%
where $\widehat{e}_{1}=\widehat{x}-x_{d}$, $\widehat{e}_{2}=\widehat{\dot{x}}%
-\dot{x}_{d}$, $\widehat{e}_{3}=\widehat{y}-y_{d}$, $\widehat{e}_{4}=%
\widehat{\dot{y}}-\dot{y}_{d}$, $\widehat{e}_{5}=\widehat{z}-z_{d}$, $%
\widehat{e}_{6}=\widehat{\dot{z}}-\dot{z}_{d}$; $k_{p1},k_{p2}>0$, and

\begin{equation}
\widehat{e}_{p}=\left[
\begin{array}{c}
\widehat{e}_{1} \\
\widehat{e}_{3} \\
\widehat{e}_{5}%
\end{array}%
\right] ,\widehat{\dot{e}}_{p}=\left[
\begin{array}{c}
\widehat{e}_{2} \\
\widehat{e}_{4} \\
\widehat{e}_{6}%
\end{array}%
\right] ,\widehat{\delta }_{p}=\left[
\begin{array}{c}
x_{1,3} \\
x_{2,3} \\
x_{3,3}%
\end{array}%
\right]
\end{equation}%
then the position error dynamics (52) rendering by controller (56) will
converge asymptotically to the origin, i.e., the tracking errors $%
e_{p}\rightarrow 0$ and $\dot{e}_{p}\rightarrow 0$ as $t\rightarrow \infty $.

The proof of Theorem 2 is presented in Appendix.

From (41), (42) and (56), we obtain

\begin{equation}
F=\left\Vert -\Xi _{p}-\widehat{\delta }_{p}-m(k_{p1}\widehat{e}_{p}+k_{p2}%
\widehat{\dot{e}}_{p})\right\Vert _{2}
\end{equation}

\newpage

\emph{4.2.2 Controller design for attitude dynamics}

\emph{Theorem 3:} For the attitude dynamics (40), to track the reference
attitude ($\psi _{d},\theta _{d},\phi _{d}$), if the controller is selected
as

\begin{equation}
u_{a}=-\Xi _{a}-\widehat{\delta }_{a}-J(k_{a1}\widehat{e}_{a}+k_{a2}\widehat{%
\dot{e}}_{a})
\end{equation}%
where $\widehat{e}_{7}=\widehat{\psi }-\psi _{d}$, $\widehat{e}_{8}=\widehat{%
\dot{\psi}}-\dot{\psi}_{d}$, $\widehat{e}_{9}=\widehat{\theta }-\theta _{d}$%
, $\widehat{e}_{10}=\widehat{\dot{\theta}}-\dot{\theta}_{d}$, $\widehat{e}%
_{11}=\widehat{\phi }-\phi _{d}$, $\widehat{e}_{12}=\widehat{\dot{\phi}}-%
\dot{\phi}_{d}$; $k_{a1},k_{a2}>0$, and

\begin{equation}
\widehat{e}_{a}=\left[
\begin{array}{c}
\widehat{e}_{7} \\
\widehat{e}_{9} \\
\widehat{e}_{11}%
\end{array}%
\right] ,\widehat{\dot{e}}_{a}=\left[
\begin{array}{c}
\widehat{e}_{8} \\
\widehat{e}_{10} \\
\widehat{e}_{12}%
\end{array}%
\right] ,\widehat{\delta }_{a}=\left[
\begin{array}{c}
x_{4,3} \\
x_{5,3} \\
x_{6,3}%
\end{array}%
\right]
\end{equation}%
then the attitude error dynamics (54) rendering by controller (59) will
converge asymptotically to the origin, i.e., the tracking errors $%
e_{a}\rightarrow 0$ and $\dot{e}_{a}\rightarrow 0$ as $t\rightarrow \infty $.

The proof of Theorem 3 is presented in Appendix.

\bigskip

\emph{4.3 Computational analysis and simulation on quadrotor aircraft}

In this section, we use a simulation on a quadrotor aircraft to illustrate
the effectiveness of the proposed estimate and control methods. When the
quadrotor desired attitude is calculated to track the translational
trajectories, the under-actuated dynamics nature exists. Here, only the
performance of the proposed observers are validated, and the reference
trajectory is selected to make the desired attitude satisfy $(\psi
_{d},\theta _{d},\phi _{d})=(0,0,0)$. The goal is to force the aircraft to
track a reference trajectory in the vertical direction. Here, the quadrotor
aircraft tracks a given trajectory ($x_{d},y_{d},z_{d}$) without the
information of ($\dot{x}$, $\dot{y}$, $\dot{z}$, $\psi $, $\theta $, $\phi $%
, $d_{1}$, $d_{2}$, $d_{3}$, $d_{4}$, $d_{5}$, $d_{6}$).

The observers (48) and (50) are used to estimate ($\dot{x}$, $\dot{y}$, $%
\dot{z}$, $\psi $, $\theta $, $\phi $, $d_{1}$, $d_{2}$, $d_{3}$, $d_{4}$, $%
d_{5}$, $d_{6}$) from the measurements of position ($x,y,z$) and the angular
velocity ($\dot{\psi},\dot{\theta},\dot{\phi}$). The controllers (56) and
(59) are presented to stabilize the flight dynamics. On the other hand, the
estimation performances by the observer (50) are compared with those by the
Extended Kalman filter [23].

Here, the aircraft is driven to move from $(0,0,0)$ to $(0,0,h_{z})$. The
reference trajectory is arranged as the following expression [29]:

\begin{equation*}
x_{d}=0,\dot{x}_{d}=0,\ddot{x}_{d}=0;y_{d}=0,\dot{y}_{d}=0,\ddot{y}_{d}=0;
\end{equation*}

\begin{equation*}
z_{d}=h_{0}(1-e^{-0.5k_{m}at^{2}}),\dot{z}%
_{d}=h_{0}k_{m}ate^{-0.5k_{m}at^{2}},\ddot{z}%
_{d}=h_{0}k_{m}a(1-k_{m}at^{2})e^{-0.5k_{m}at^{2}}
\end{equation*}

The initial states of quadrotor aircraft are: ($x(0)$, $\dot{x}(0)$, $y(0)$,
$\dot{y}(0)$, $z(0)$, $\dot{z}(0)$, $\psi (0)$, $\dot{\psi}(0)$, $\theta (0)$%
, $\dot{\theta}(0)$, $\phi (0)$, $\dot{\phi}(0)$) = ($0.5$, $-0.5$, $-0.5$, $%
0.5$, $0.5$, $-1$, $0.2$, $0.3$, $0.3$, $-0.1$, $0.2$, $-0.2$); the initial
states of the observers are: ($x_{1,1}(0)$, $x_{1,2}(0)$, $x_{1,3}(0)$, $%
x_{2,1}(0)$, $x_{2,2}(0)$, $x_{2,3}(0)$, $x_{3,1}(0)$, $x_{3,2}(0)$, $%
x_{3,3}(0)$, $x_{4,1}(0)$, $x_{4,2}(0)$, $x_{4,3}(0)$, $x_{5,1}(0)$, $%
x_{5,2}(0)$, $x_{5,3}(0)$, $x_{6,1}(0)$, $x_{6,2}(0)$, $x_{6,3}(0)$) = ($0$ $%
0$, $0$, $0$, $0$, $0$, $0$, $0$, $0$, $0.2$, $0.3$, $0$, $0.3$, $-0.1$, $0$%
, $0.2$, $-0.2$, $0$). Let the uncertainties be: $\delta _{x}=0.5\sin (t)$, $%
\delta _{y}=0.5\sin (t)$, $\delta _{z}=0.5\sin (t)$, $\delta _{\psi
}=0.2\sin (0.8t)$, $\delta _{\theta }=0.2\sin (0.8t)$, $\delta _{\phi
}=0.2\sin (0.8t)$.

\begin{figure}[H]
\begin{center}
\includegraphics[width=2.90in]{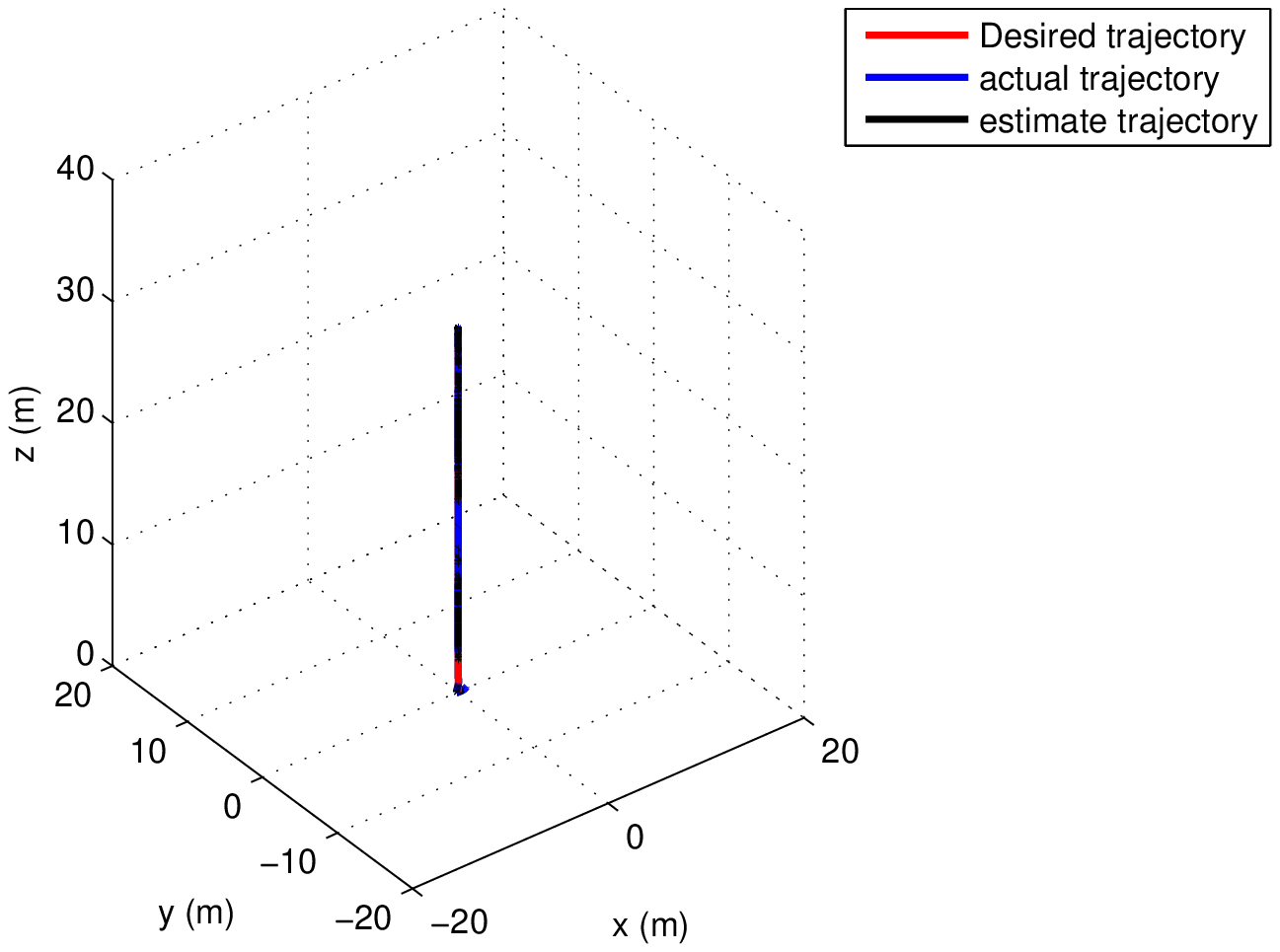}\\[0pt]
{8(a) Position trajectory}\\[0pt]
\includegraphics[width=2.90in]{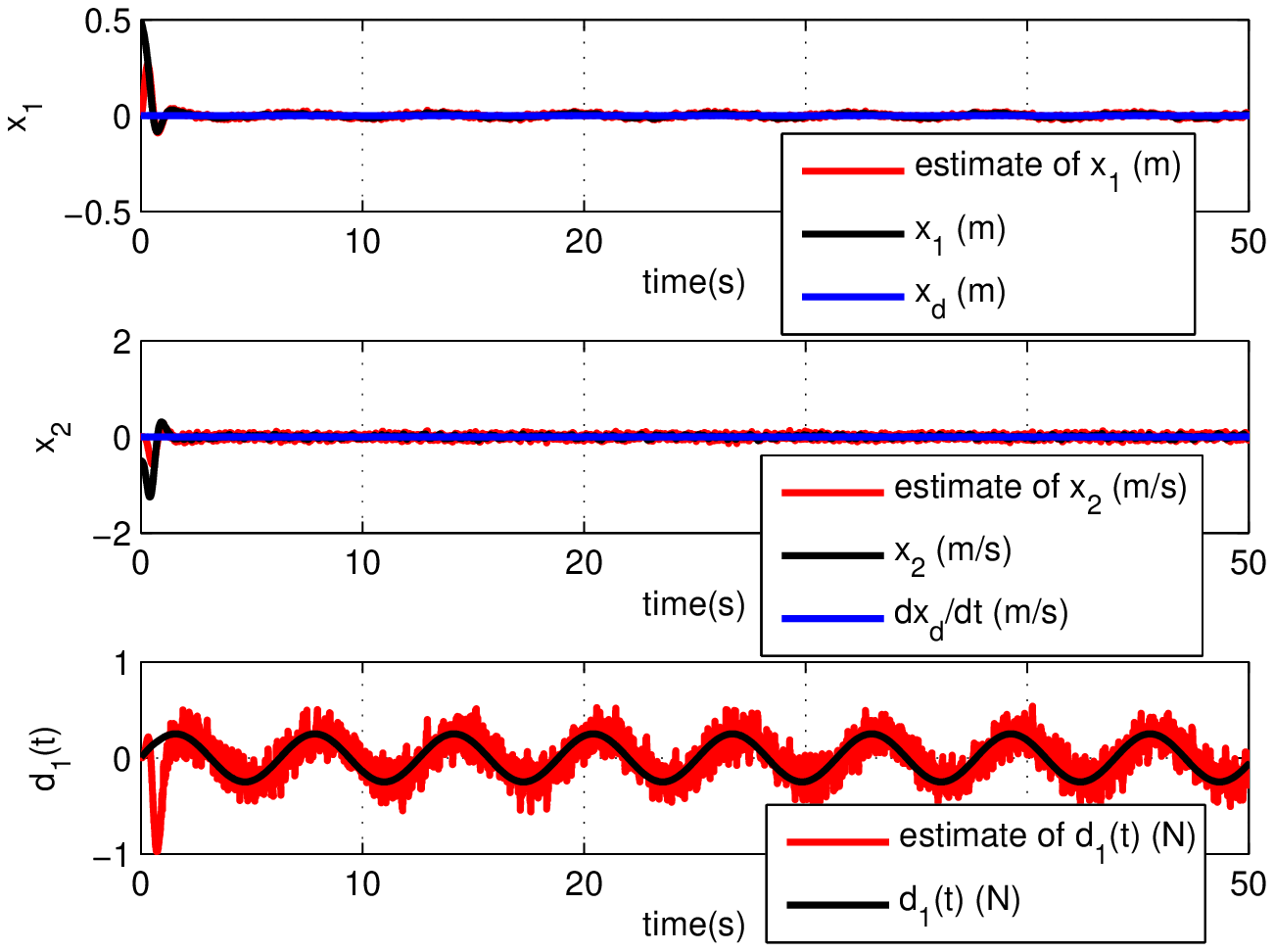}\\[0pt]
{8(b) Estimation in x coordinate\\[0pt]
\includegraphics[width=2.90in]{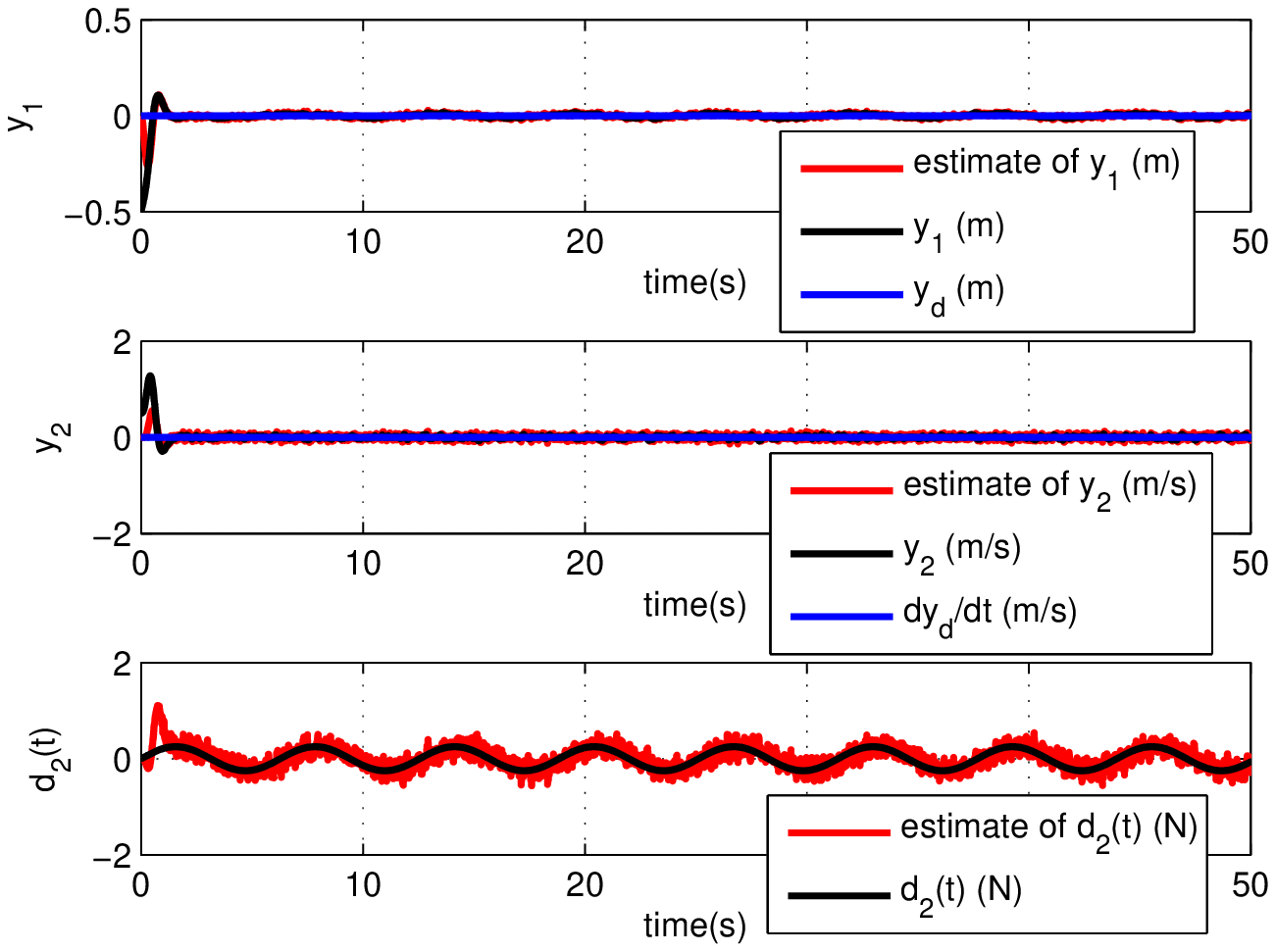}\\[0pt]
8(c) Estimation in y coordinate\\[0pt]
\includegraphics[width=2.90in]{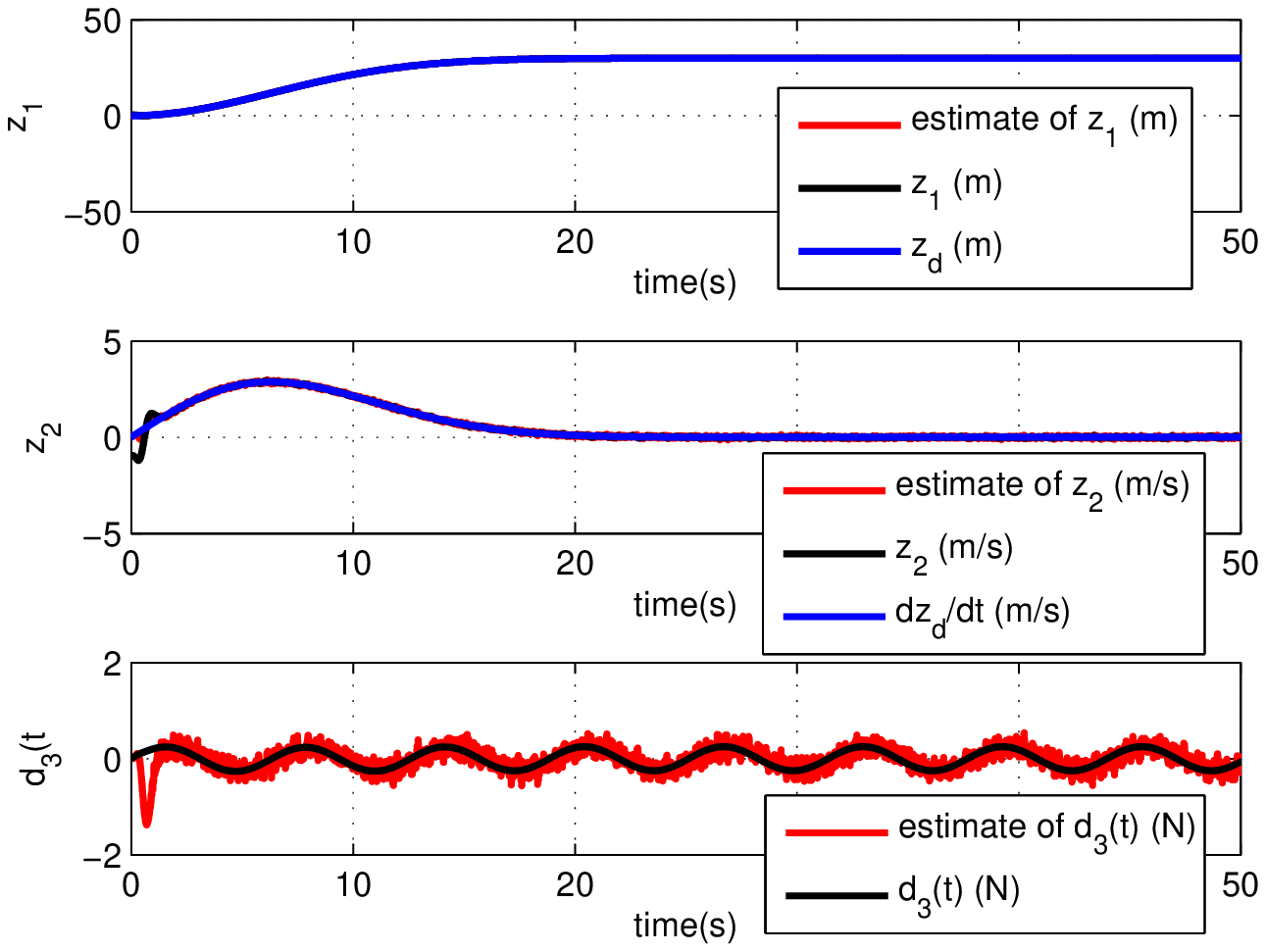}\\[0pt]
8(d) Estimation in z coordinate}
\end{center}
\end{figure}

\begin{figure}[H]
\begin{center}
\includegraphics[width=2.90in]{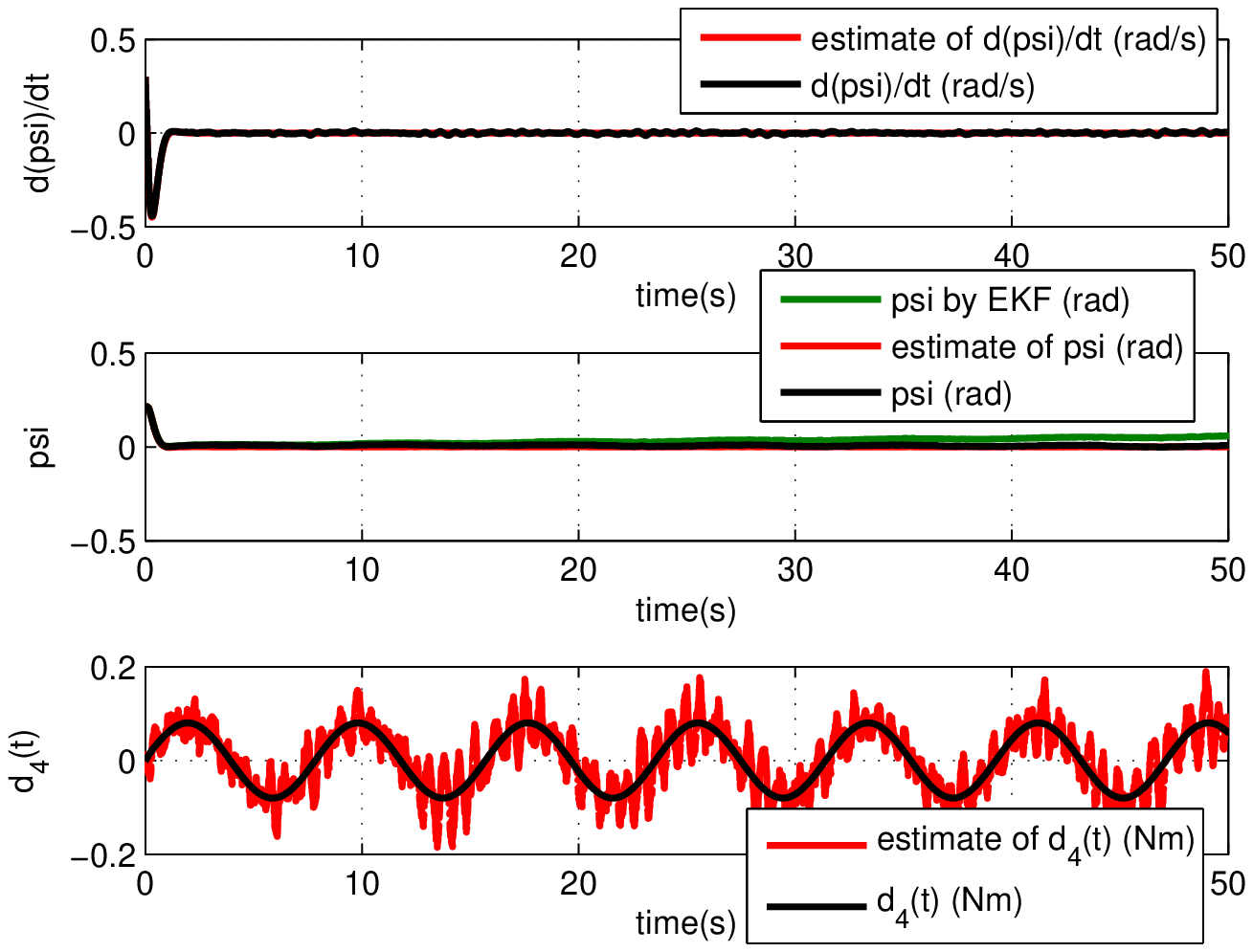}\\[0pt]
{8(e) Estimation in yaw dynamics}\\[0pt]
\includegraphics[width=2.90in]{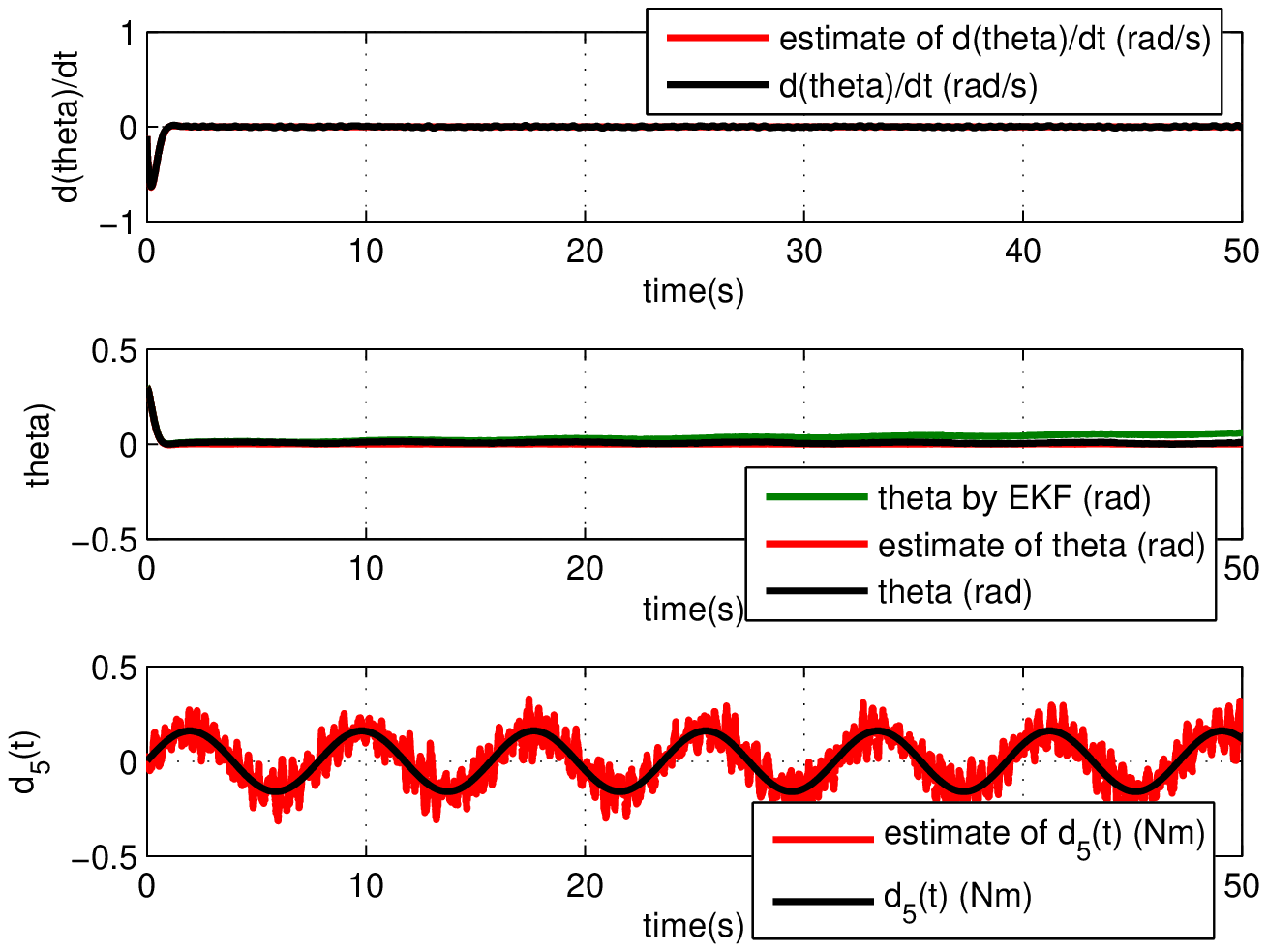}\\[0pt]
{8(f) Estimation in pitch dynamics\\[0pt]
\includegraphics[width=2.90in]{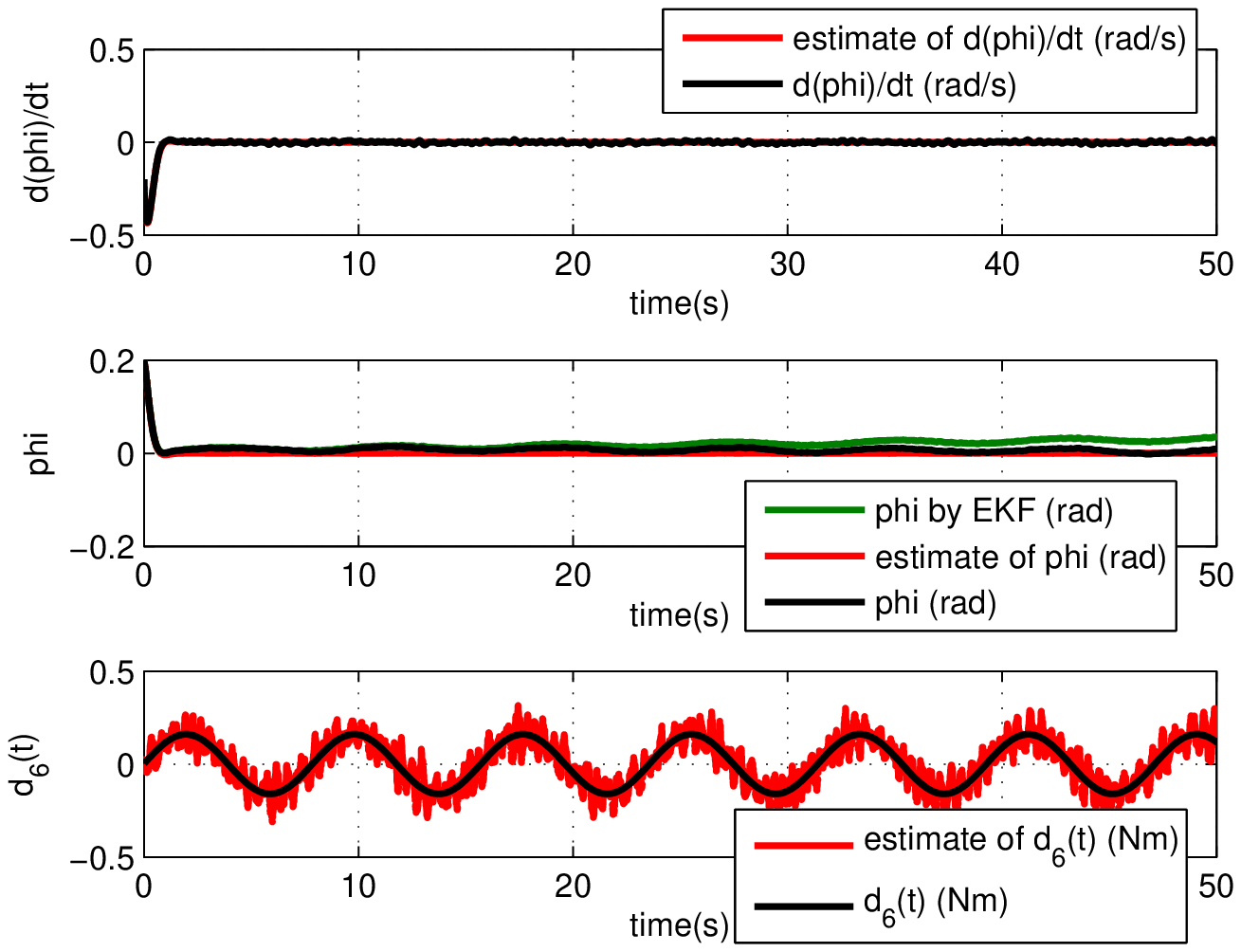}\\[0pt]
8(g) Estimation in roll dynamics\\[0pt]
\includegraphics[width=2.90in]{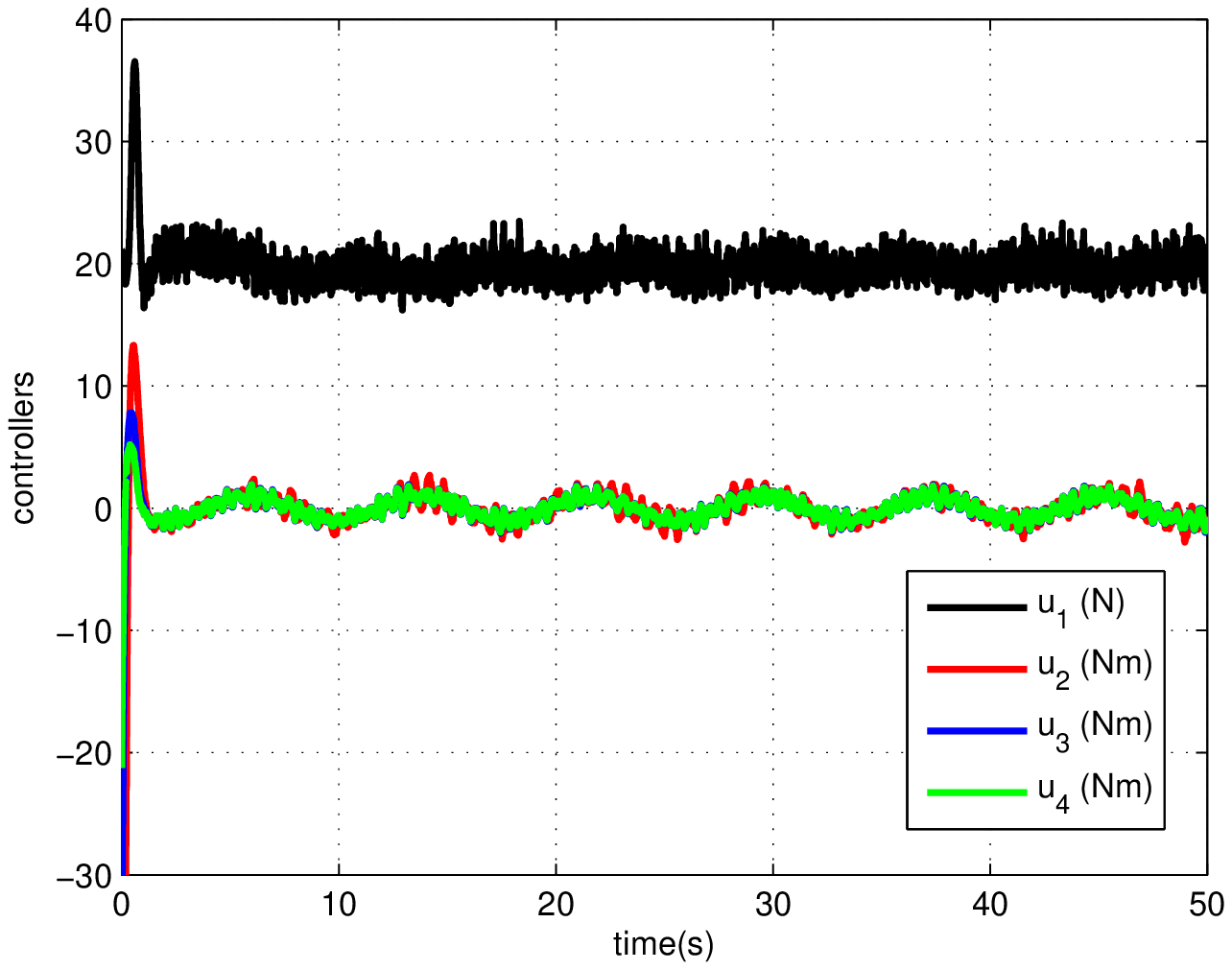}\\[0pt]
8(h) Controller}\\[0pt]
{Figure 8 Quadrotor aircraft control based on differentiation-integration
observer in 50s}
\end{center}
\end{figure}

\begin{figure}[H]
\begin{center}
\includegraphics[width=2.90in]{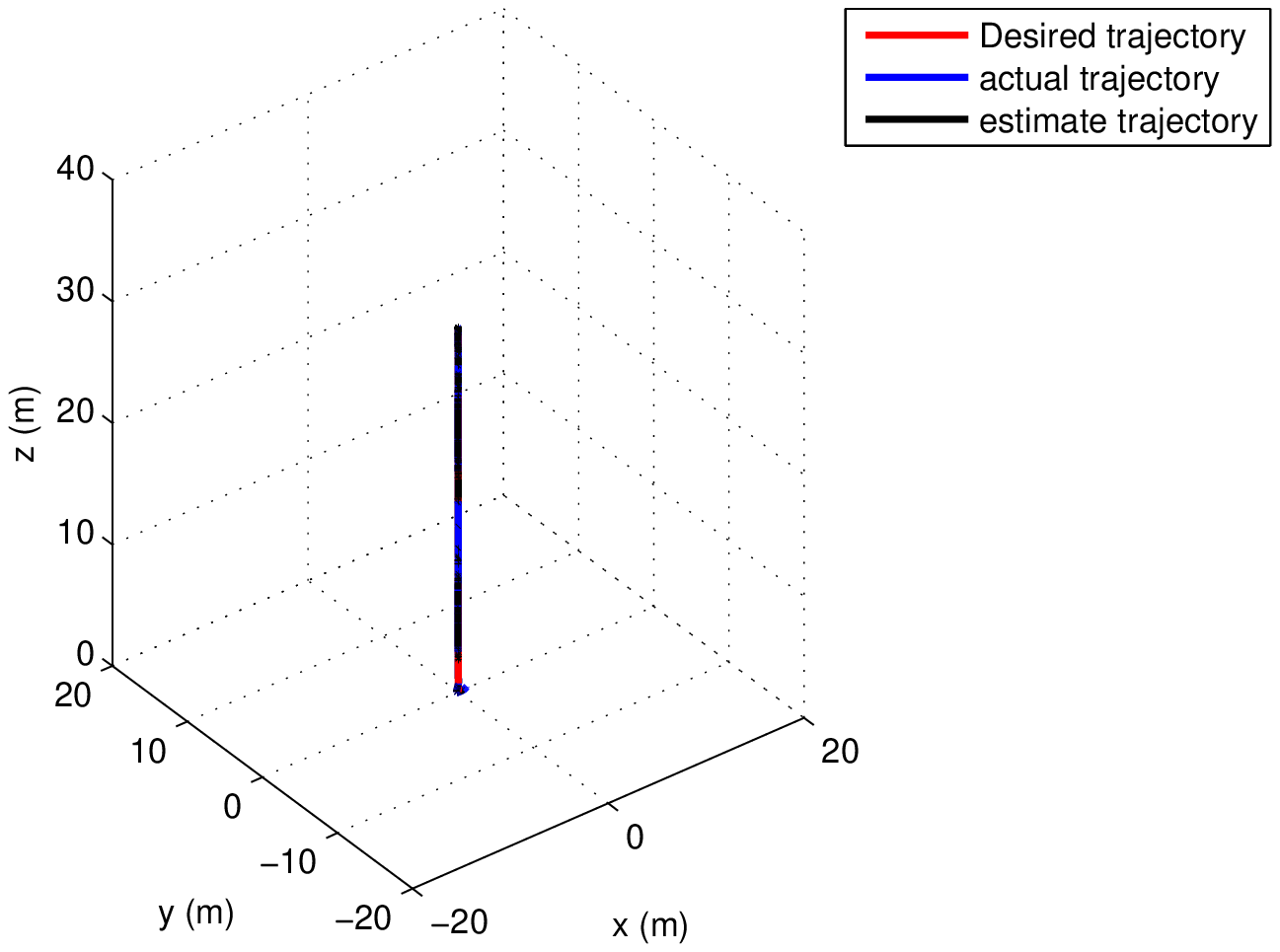}\\[0pt]
{9(a) Position trajectory}\\[0pt]
\includegraphics[width=2.90in]{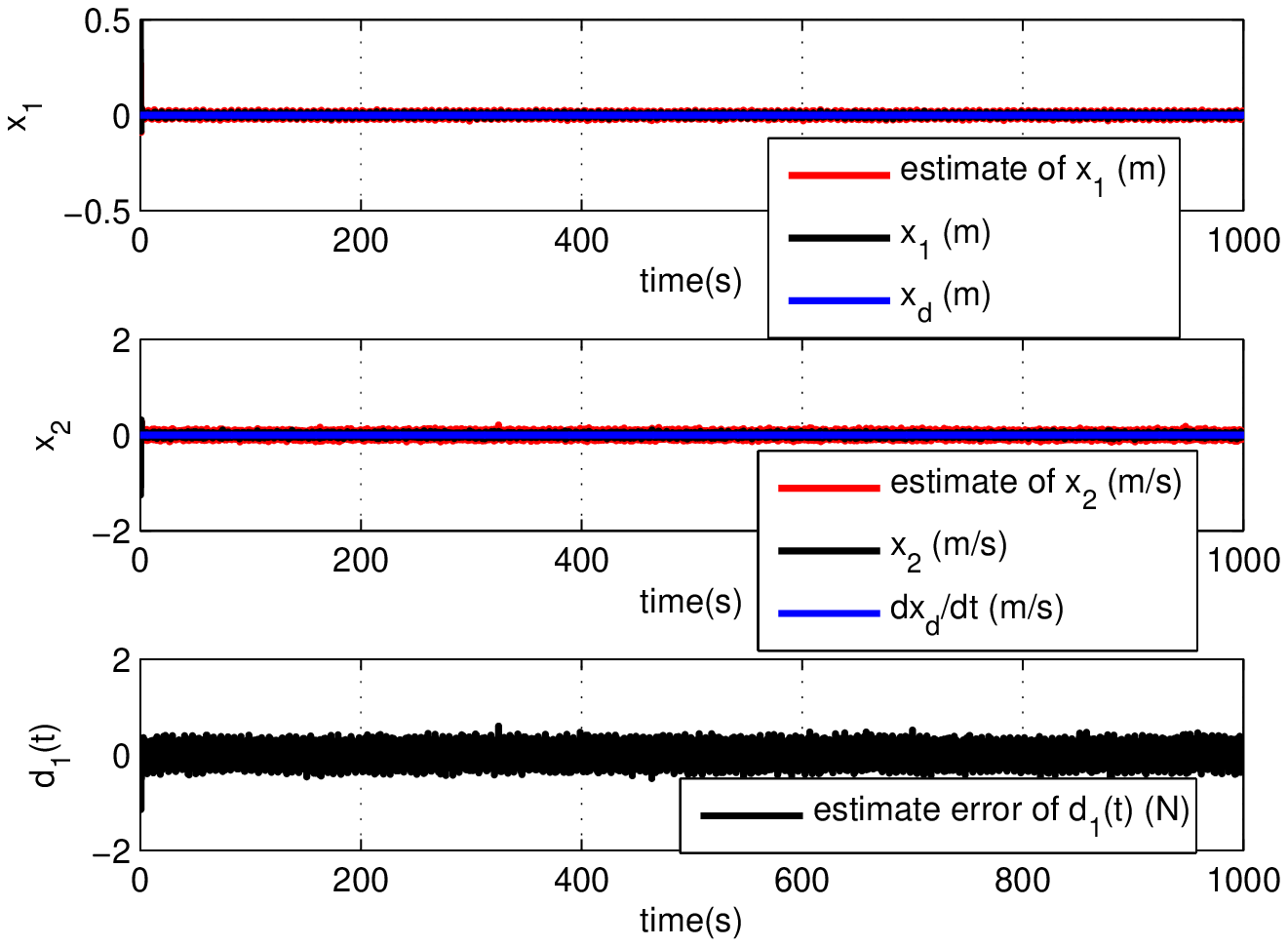}\\[0pt]
{9(b) Estimation in x coordinate\\[0pt]
\includegraphics[width=2.90in]{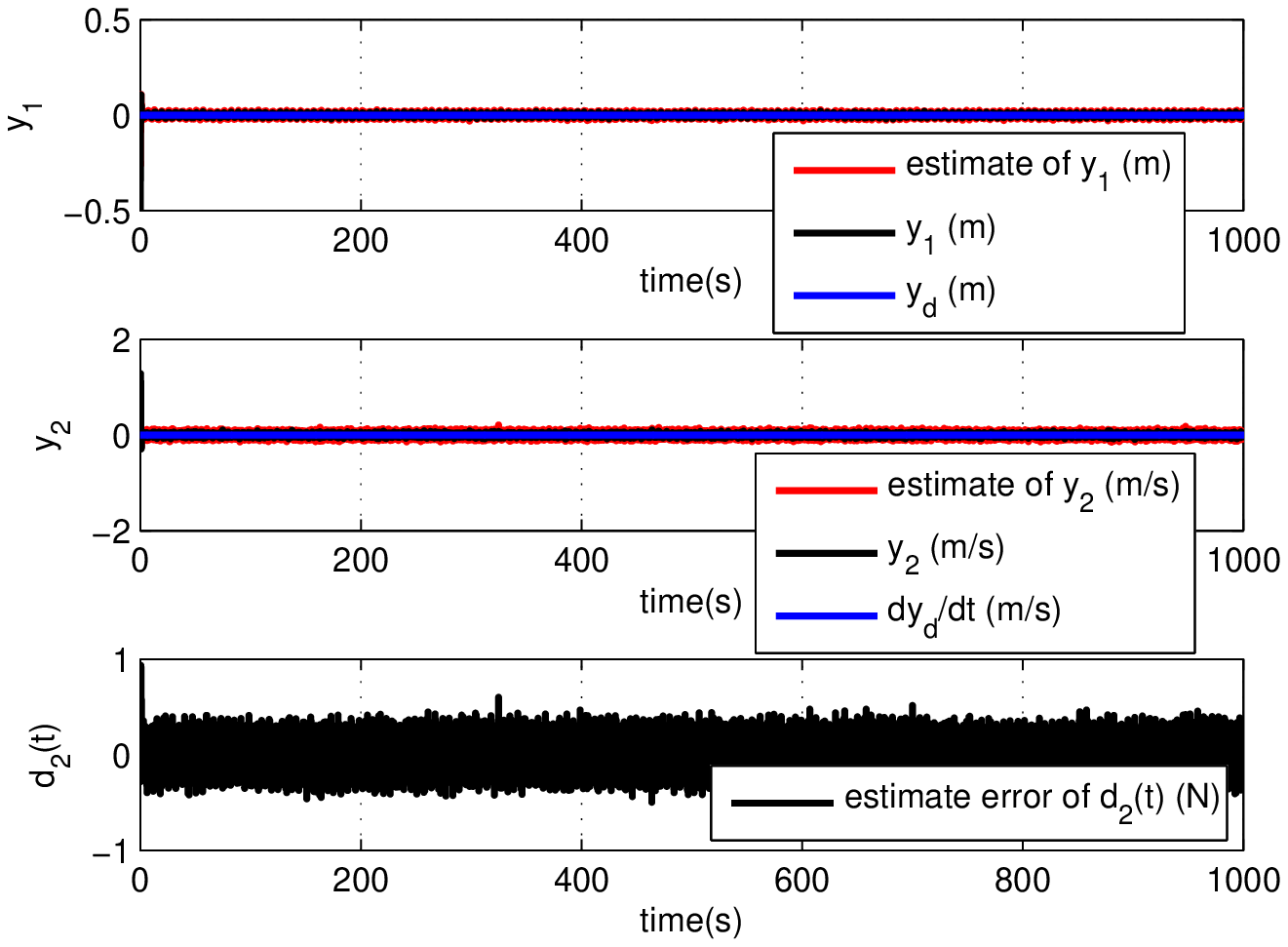}\\[0pt]
9(c) Estimation in y coordinate\\[0pt]
\includegraphics[width=2.90in]{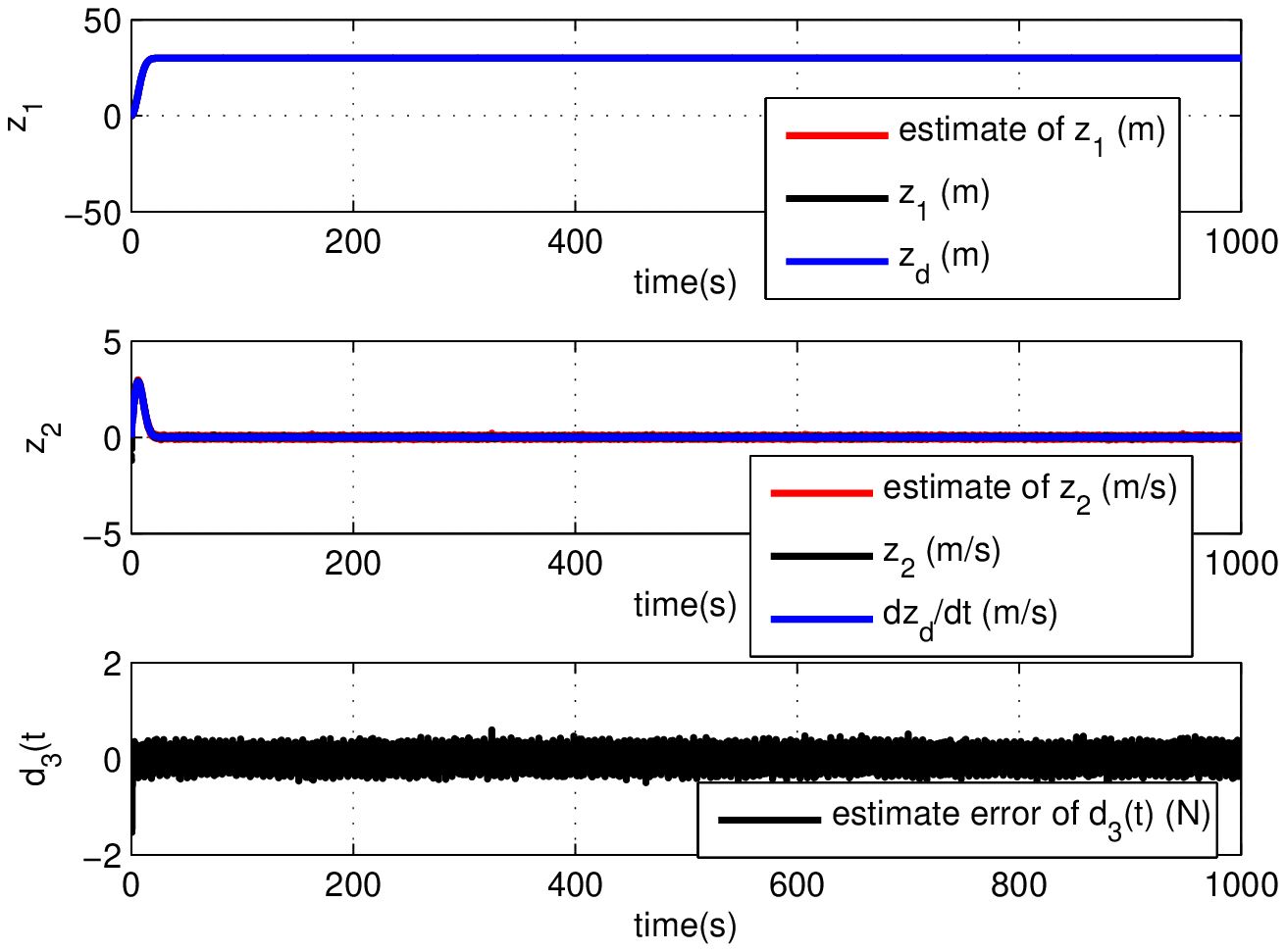}\\[0pt]
9(d) Estimation in z coordinate}
\end{center}
\end{figure}

\begin{figure}[H]
\begin{center}
\includegraphics[width=2.90in]{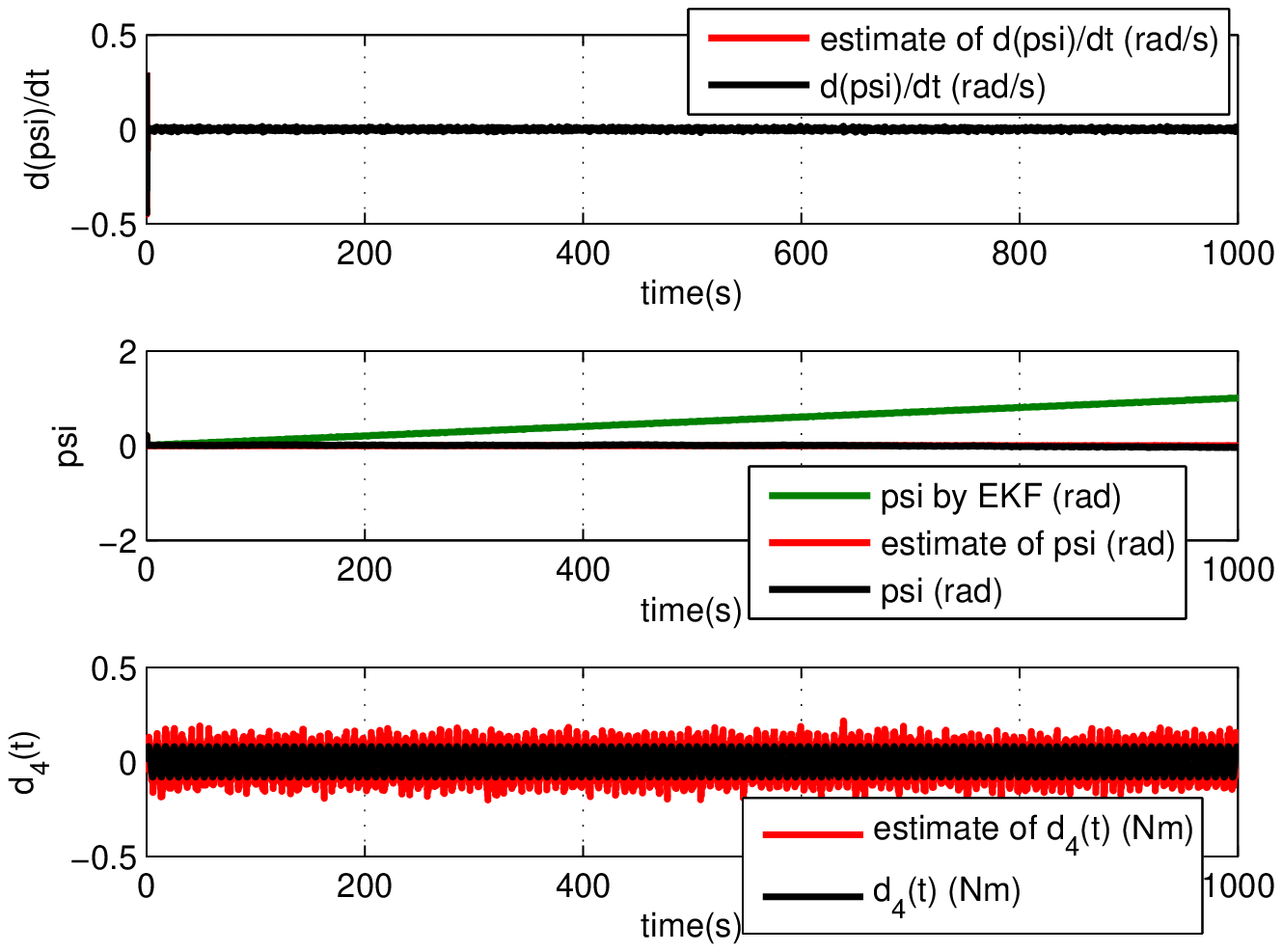}\\[0pt]
{9(e) Estimation in yaw dynamics}\\[0pt]
\includegraphics[width=2.90in]{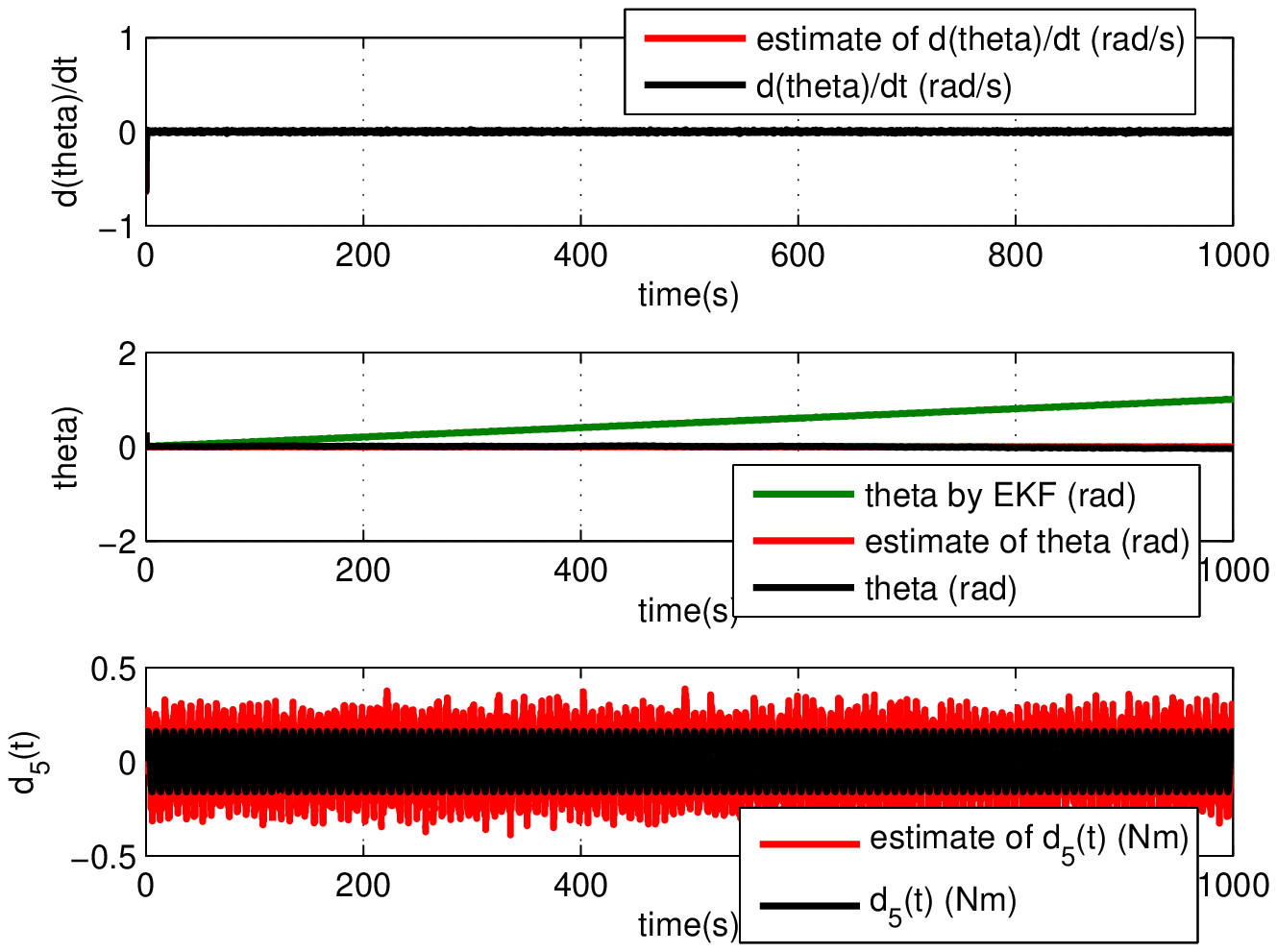}\\[0pt]
{9(f) Estimation in pitch dynamics\\[0pt]
\includegraphics[width=2.90in]{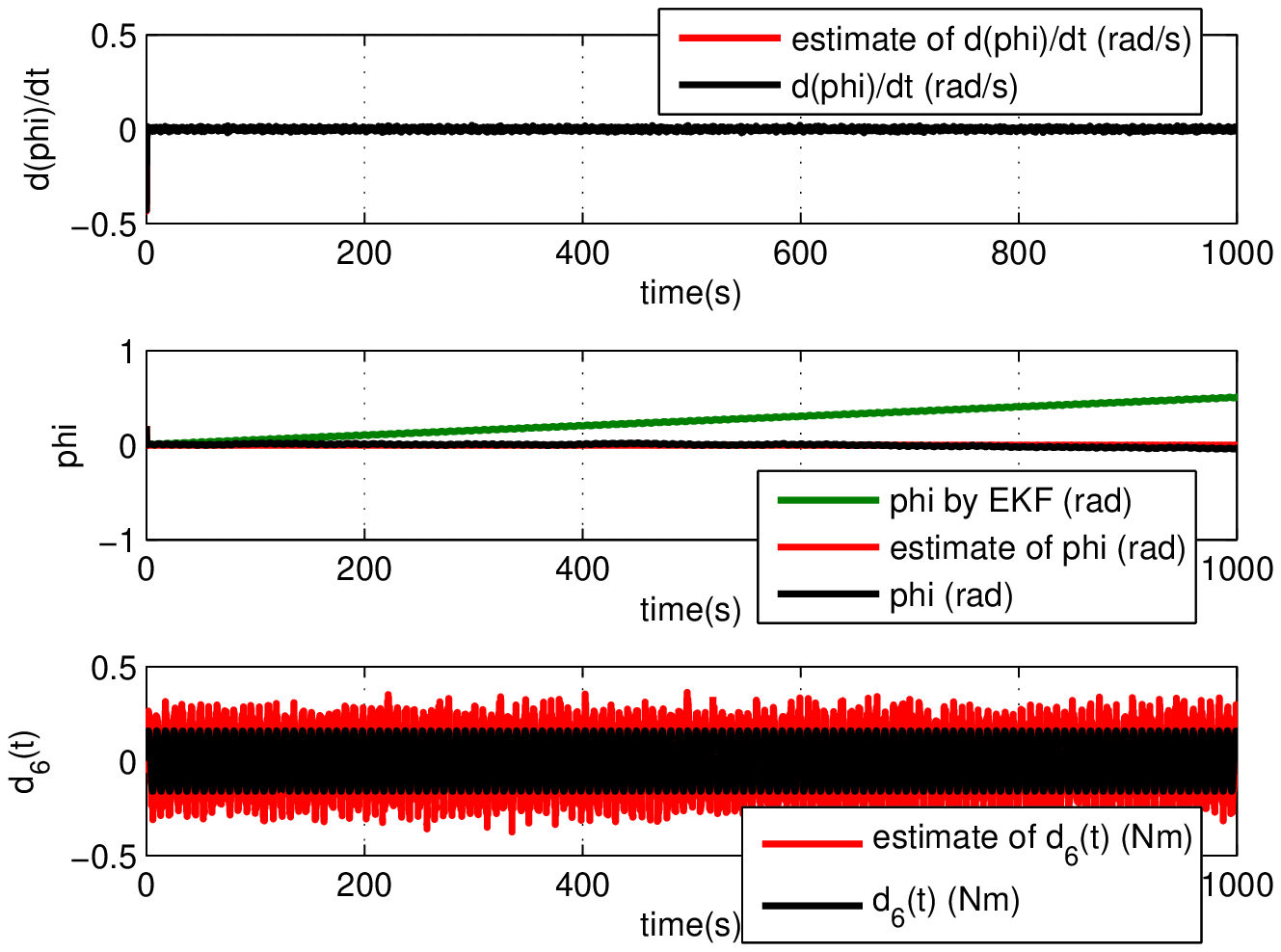}\\[0pt]
9(g) Estimation in roll dynamics\\[0pt]
\includegraphics[width=2.90in]{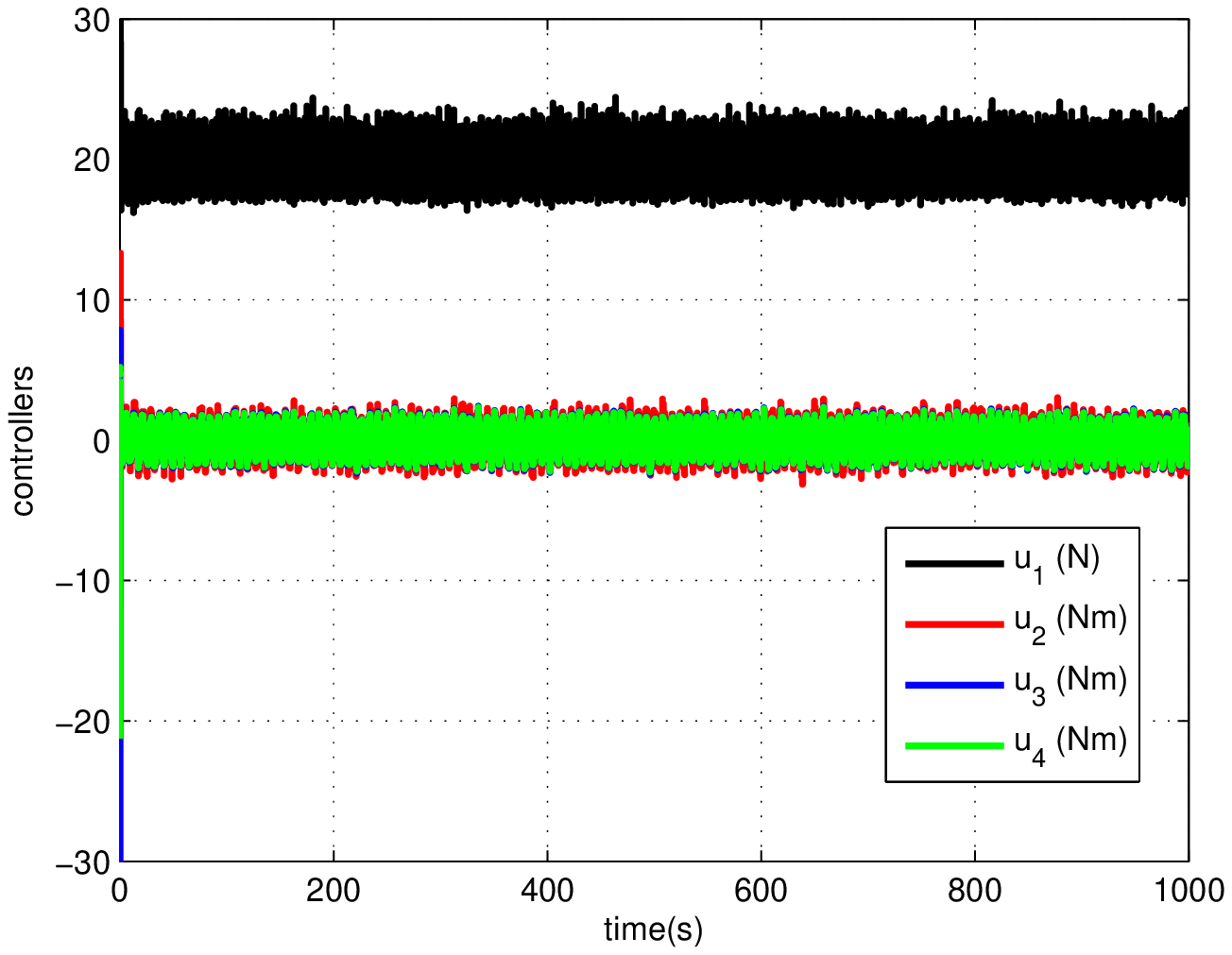}\\[0pt]
9(h) Controller}\\[0pt]
{Figure 9 Quadrotor aircraft control based on differentiation-integration
observer in 1000s}
\end{center}
\end{figure}

The position measurement outputs are $y_{opi}=w_{i,1}+n_{i}$, where $i=1,2,3$%
, and $w_{1,1}=x$, $w_{2,1}=y$, $w_{3,1}=z$; the attitude measurement
outputs are $y_{opi}=w_{i,2}+n_{i}$, where $i=4,5,6$, and $w_{4,2}=\dot{\psi}
$, $w_{5,2}=\dot{\theta}$, $w_{6,2}=\dot{\phi}$; $n_{i}$ (where $i=1,\cdots
,6$) are the disturbances.

The disturbances $n_{i}$ (where $i=1,\cdots ,6$) include two types of
noises: Random number with Mean=0, Variance=0.001, Initial speed=0, and
Sample time=0; Pulses with Amplitude=0.001, Period=1s, Pulse width=1, and
Phase delay=0.

The parameters of the aircraft control system are given as follows:

\emph{Quadrotor aircraft [29]:} $m=2kg$, $g=9.81m/s^{2}$, $l=0.2m$, $%
J_{x}=1.25Ns^{2}/rad$, $J_{y}=1.25Ns^{2}/rad$, $J_{z}=2.5Nm$; $b=2.923\times
10^{-3}$, $k=5\times 10^{-4}$, $k_{x}=k_{y}=k_{z}=0.01Ns/m$, $k_{\psi
}=k_{\theta }=k_{\phi }=0.012Ns/rad$;

\emph{Third-order differentiator:} $k_{i,1}=6$, $k_{i,2}=11$, $k_{i,3}=6$, $%
i=1,2,3$;

In order to reduce the peaking phenomena in the outputs of the
differentiator due to the large initial observation errors, the perturbation
parameters are selected as $1/\varepsilon _{i}=\left\{
\begin{array}{c}
5t,t\leq 1 \\
5,t>1%
\end{array}%
\right. ,i=1,2,3$;

\emph{Differentiation-integration observer:} $k_{2,i,1}=0.1$, $k_{2,i,2}=2$,
$k_{2,i,3}=1$, $i=4,5,6$;

Because the initial observation errors are small for the
differentiation-integration observer, no chattering phenomenon happen, and
the perturbation parameter can be selected as $\varepsilon _{i}=1/3$, $%
i=4,5,6$;

\emph{Reference trajectory:} $h_{0}=30m$, $a=5m/s^{2}$, $k_{m}=0.005$;

\emph{Controllers:} $k_{p1}=16$, $k_{p2}=8$, $k_{a1}=28$, $k_{a2}=8$.

In this simulation, without the information of velocity, attitude angle and
uncertainties, the aircraft is controlled to track the reference trajectory.
The position is obtained from the GPS receiver, and the altitude information
is from the altimeter. The angular velocity ($\dot{\psi},\dot{\theta},\dot{%
\phi}$) is measured by the IMU. Differentiation-integration observer (50) is
used to estimate the attitude angle ($\psi ,\theta ,\phi $) and
uncertainties in the attitude dynamics from measurement of the angular
velocity ($\dot{\psi},\dot{\theta},\dot{\phi}$). The third-order
differentiator (48) is adopted to estimate the velocity ($\dot{x},\dot{y},%
\dot{z}$) and uncertainties in the position dynamics from the measurement of
position ($x,y,z$). Controllers (56) and (59) are designed to control the
aircraft to track the reference trajectory.

The data of flying test are presented in Figs.8 and 9. Fig. 8(a) describes
the position trajectory; Fig. 8(b) describes the estimate of $x$, $dx/dt$
and $d_{1}(t)$; Fig. 8(c) describes the estimate of $y$, $dy/dt$ and $%
d_{2}(t)$; Fig. 8(d) presents the estimate of $z$, $dz/dt$ and $d_{3}(t)$;
Fig. 8(e) presents the estimate of the yaw angle $\psi $, yaw rate $d\psi
/dt $ and $d_{4}(t)$; Fig. 8(f) presents the estimate of the pitch angle $%
\theta $, pitch rate $d\psi /dt$ and $d_{5}(t)$; Fig. 8(g) presents the
estimate of the roll angle $\phi $, roll rate $d\phi /dt$ and $d_{1}(t)$;
Fig. 8(h) presents the controller curves of $u_{1}$, $u_{2}$, $u_{3}$ and $%
u_{4}$. The simulation in 1000s is proposed in Fig. 9. In the simulation
above, though high-frequency stochastic noises exist in the measurement
signals, the uncertainties exist in the aircraft dynamics, and only the
angular velocity is considered in the IMU output, the attitude estimations
by the presented differentiation-integration observer, the
velocity-uncertainty estimations by the third-order differentiator and the
control results by the designed controller have satisfying qualities. The
stochastic noises are restrained sufficiently by the differentiator and
differentiation-integration observer. Furthermore, from Fig. 9, no
chattering and drift phenomena happen for the differentiation-integration
observer in long-time simulation. In the tracking outputs, not only the
dynamical performances are fast, but also the tracking precisions are
accurate. Importantly, due to the satisfying estimate precision and the
strong robustness of the observers, a very simple control law can be
selected to implement the satisfying tracking qualities. However, from Figs.
8(e), 8(f), 8(g), 9(e), 9(f) and 9(g), the obvious position drifts exist in
the outputs of the integral algorithm by the Extended Kalman filter. The
integral algorithm can't restrain the effect of stochastic noise (especially
non-white noise). Such noise leads to the accumulation of additional drift
in the integrated signal.

\section{Conclusions}

In this paper, a generalized differentiation-integration observer has been
developed, which can estimate the integrals and derivatives of a signal,
synchronously. The proposed observers have built in low pass filters. It can
achieve better performance without additional filters. The effectiveness of
the proposed differentiation-integration observer was shown by the
simulations: i) it succeeded in estimating the integrals and derivatives of
the measurement signal; ii) Due to the satisfying estimate precision and the
strong robustness, the differentiation-integration observers are suitable to
the controller design for quadrotor aircraft. The merits of the presented
differentiation-integration observer include its synchronous estimation of
integrals and derivatives, simple implementation, sufficient stochastic
noises rejection and almost no drift phenomenon. Although high-frequency
stochastic noises and measurement errors exist, the integral and derivative
estimations by the proposed observer and the tracking results by the
designed controller for the quadrotor aircraft have the satisfying qualities.

ACKNOWLEDGEMENT

This work is partially supported by Research Grants Council, Hong Kong, SAR
PR China under RGC16200514.

\bigskip

\begin{center}
{\Large Appendix}
\end{center}

\emph{Proof of Lemma 1:} Let $\overline{k}_{i}=k_{i}$, and $\overline{k}_{p}=%
\frac{k_{p}}{\varepsilon ^{p-c(p)}}$, where $p\neq i$, $i=1,\cdots ,n$. Then
the polynomial (5) can be rewritten as

\begin{equation}
s^{n}+\overline{k}_{n}s^{n-1}+\cdots +\overline{k}_{p}s^{p-1}+\cdots +%
\overline{k}_{2}s+\overline{k}_{1}
\end{equation}

The Routh table of the polynomial (61) is presented in Fig. 10. We found
that the Routh table is the nested structure.

1) when $n\in \{1,2,\cdots \}$ and $p=1$, we know that $\overline{k}%
_{i}=k_{i}$, where $i=1,\cdots ,n$. From the Routh table in figure 10, the
parameters $k_{i}>0$ (where $i=1,\cdots ,n$) can be selected such that the
polynomial $s^{n}+\sum\limits_{{i=1}}^{{n}}k_{i}s^{i-1}$ is Hurwitz.

If there exists an integer $N$, such that, when $n=N$ and $p\in \{2,3,\cdots
,n\}$, for the arbitrary large $\overline{k}_{p}$, the polynomial (61)
cannot be Hurwitz, then, from the nested structure of the Routh table, when $%
n\geq N$ and $p\in \{2,3,\cdots ,n\}$, for the arbitrary large $\overline{k}%
_{p}$, the polynomial (61) cannot be Hurwitz.

\begin{figure}[H]
\begin{center}
\includegraphics[width=6.00in]{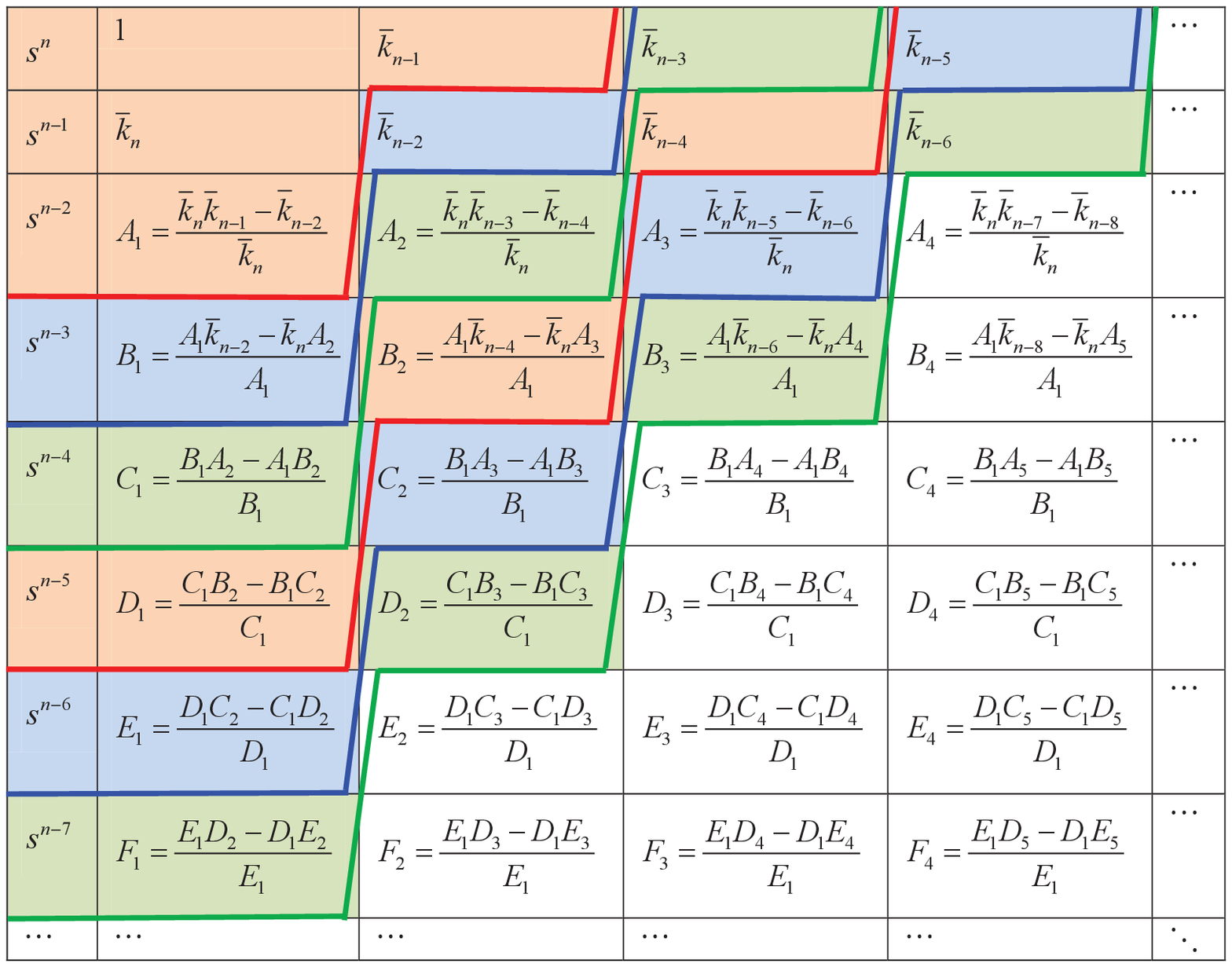}\\[0pt]
{\small Figure 10 The Routh table of the polynomial (61)}
\end{center}
\end{figure}

2) when $n=2$, from the Routh table in Fig. 10, the polynomial $s^{2}+%
\overline{k}_{2}s+\overline{k}_{1}$ is Hurwitz if $\overline{k}_{2}=\frac{%
k_{2}}{\varepsilon ^{2-c(2)}}>0$ and $\overline{k}_{1}=k_{1}>0$. That is to
say, for the arbitrary $\varepsilon \in \left( 0,1\right) $, the polynomial $%
s^{2}+\frac{k_{2}}{\varepsilon ^{2-c(2)}}s+k_{1}$ is Hurwitz if $k_{1}>0$
and $k_{2}>0$.

3) when $n=3$, from the Routh table in Fig. 10, the polynomial $s^{3}+%
\overline{k}_{3}s^{2}+\overline{k}_{2}s+\overline{k}_{1}$ is Hurwitz if $%
\overline{k}_{3}>0$, $A_{1}>0$ and $B_{1}>0$, i.e, $\overline{k}_{3}>0$, $%
\overline{k}_{1}>0$, and $\overline{k}_{3}\overline{k}_{2}>\overline{k}_{1}$%
. Therefore: i) when $p=2$ (i.e., $\overline{k}_{2}=\frac{k_{2}}{\varepsilon
^{2-c(2)}}$), for the arbitrary $\varepsilon \in \left( 0,1\right) $,
polynomial $s^{3}+k_{3}s^{2}+\frac{k_{2}}{\varepsilon ^{2-c(p)}}s+k_{1}$ is
Hurwitz if $k_{1}>0$, $k_{3}>0$, $k_{2}>\varepsilon ^{2-c(p)}\frac{k_{1}}{%
k_{3}}$; ii) when when $p=3$ (i.e., $\overline{k}_{3}=\frac{k_{3}}{%
\varepsilon ^{3-c(3)}}$), for the arbitrary $\varepsilon \in \left(
0,1\right) $, polynomial $s^{3}+\frac{k_{3}}{\varepsilon ^{3-c(3)}}%
s^{2}+k_{2}s+k_{1}$ is Hurwitz if $k_{1}>0$, $k_{3}>0$, $k_{2}>\varepsilon
^{3-c(3)}\frac{k_{1}}{k_{3}}$.

4) when $n=4$, from the Routh table in Fig. 10, the polynomial $s^{4}+%
\overline{k}_{4}s^{3}+\overline{k}_{3}s^{2}+\overline{k}_{2}s+\overline{k}%
_{1}$ is Hurwitz if $\overline{k}_{4}>0$, $A_{1}>0$, $B_{1}>0$, and $C_{1}>0$%
, i.e, $\overline{k}_{4}>0$, $\overline{k}_{4}\overline{k}_{3}>\overline{k}%
_{2}$, $\frac{\overline{k}_{4}\overline{k}_{3}-\overline{k}_{2}}{\overline{k}%
_{4}}\overline{k}_{2}>\overline{k}_{4}\overline{k}_{1}$, $\overline{k}_{1}>0$%
. We find that, for the arbitrary $\varepsilon \in \left( 0,1\right) $, only
when $p=3$, the polynomial $s^{4}+k_{4}s^{3}+\frac{k_{3}}{\varepsilon
^{3-c(3)}}s^{2}+k_{2}s+k_{1}$ is Hurwitz if $k_{1}>0$, $k_{4}>0$, $%
k_{3}>\varepsilon ^{3-c(3)}\frac{k_{2}}{k_{4}}$, and $k_{2}>\varepsilon
^{3-c(3)}\frac{k_{4}^{2}k_{1}+k_{2}^{2}}{k_{4}k_{3}}$.

5) when $n=5$, from the Routh table in Fig. 10, the polynomial $s^{5}+%
\overline{k}_{5}s^{4}+\overline{k}_{4}s^{3}+\overline{k}_{3}s^{2}+\overline{k%
}_{2}s+\overline{k}_{1}$ is Hurwitz if $\overline{k}_{5}>0$, $A_{1}>0$, $%
B_{1}>0$, $C_{1}>0$, and $D_{1}>0$. We found that, for arbitrary large $%
\overline{k}_{p}=\frac{k_{p}}{\varepsilon ^{p}}$ and all the $p\in
\{2,3,4,5\}$, no matter how to select $\overline{k}_{i}$ (where $i=1,\cdots
,5$), the polynomial $s^{5}+\overline{k}_{5}s^{4}+\overline{k}_{4}s^{3}+%
\overline{k}_{3}s^{2}+\overline{k}_{2}s+\overline{k}_{1}$ cannot be Hurwitz.

6) Therefore, from the nested structure of the Routh table, there exists an
integer $5$, when $n\geq 5$ and $p\in \{2,3,\cdots ,n\}$, for the arbitrary
large $\overline{k}_{p}=\frac{kp}{\varepsilon ^{p}}$, the polynomial (5)
cannot be Hurwitz. This concludes the proof. $\blacksquare $

\bigskip

\emph{Proof of Theorem 2:} In the light of Corollary 1, the observation
signals is satisfied with $\underset{\varepsilon \rightarrow 0}{\lim }%
\widehat{p}_{p}=p_{p}$, $\underset{\varepsilon \rightarrow 0}{\lim }\widehat{%
\dot{p}}_{p}=\dot{p}_{p}$, $\underset{\varepsilon \rightarrow 0}{\lim }%
\widehat{\delta }_{p}=\delta _{p}$, where

\begin{equation}
p_{p}=\left[
\begin{array}{c}
x \\
y \\
z%
\end{array}%
\right] ,\dot{p}_{p}=\left[
\begin{array}{c}
\dot{x} \\
\dot{y} \\
\dot{z}%
\end{array}%
\right] ,\widehat{p}_{p}=\left[
\begin{array}{c}
\widehat{x} \\
\widehat{y} \\
\widehat{z}%
\end{array}%
\right] ,\widehat{\dot{p}}_{p}=\left[
\begin{array}{c}
\widehat{\dot{x}} \\
\widehat{\dot{y}} \\
\widehat{\dot{z}}%
\end{array}%
\right]
\end{equation}

Considering controller (56), the closed-loop error system for position
dynamics is

\begin{equation}
\ddot{e}_{p}=-k_{p1}e_{p}-k_{p2}\dot{e}_{p}-k_{p1}(\widehat{p}%
_{p}-p_{p})-k_{p2}(\widehat{\dot{p}}_{p}-\dot{p}_{p})-m^{-1}(\widehat{\delta
}_{p}-\delta _{p})
\end{equation}

For $t\geq t_{s}$ and sufficiently small $\varepsilon $, selecting the
Lyapunov function be $V_{p}=k_{p1}e_{p}^{T}e_{p}+\frac{1}{2}\dot{e}_{p}^{T}%
\dot{e}_{p}$, we can obtain that $e_{p}\rightarrow 0$ and $\dot{e}%
_{p}\rightarrow 0$ as $t\rightarrow \infty $. This concludes the proof. $%
\blacksquare $

\emph{Proof of Theorem 3:} In the light of Corollary 2, the observation
signals is satisfied with $\underset{\varepsilon \rightarrow 0}{\lim }%
\widehat{p}_{a}=p_{a}$, $\underset{\varepsilon \rightarrow 0}{\lim }\widehat{%
\dot{p}}_{a}=\dot{p}_{a}$, $\underset{\varepsilon \rightarrow 0}{\lim }%
\widehat{\delta }_{a}=\delta _{a}$, where

\begin{equation}
a_{a}=\left[
\begin{array}{c}
\psi \\
\theta \\
\phi%
\end{array}%
\right] ,\dot{a}_{a}=\left[
\begin{array}{c}
\dot{\psi} \\
\dot{\theta} \\
\dot{\phi}%
\end{array}%
\right] ,\widehat{a}_{a}=\left[
\begin{array}{c}
\widehat{\psi } \\
\widehat{\theta } \\
\widehat{\phi }%
\end{array}%
\right] ,\widehat{\dot{a}}_{a}=\left[
\begin{array}{c}
\widehat{\dot{\psi}} \\
\widehat{\dot{\theta}} \\
\widehat{\dot{\phi}}%
\end{array}%
\right]
\end{equation}

Considering controller (59), the closed-loop error system for attitude
dynamics is

\begin{equation}
\ddot{e}_{a}=-k_{a1}e_{a}-k_{a2}\dot{e}_{a}-k_{a1}(\widehat{a}%
_{a}-a_{a})-k_{a2}(\widehat{\dot{a}}_{a}-\dot{a}_{a})-J^{-1}(\widehat{\delta
}_{a}-\delta _{a})
\end{equation}

For $t\geq t_{s}$ and sufficiently small $\varepsilon $, selecting the
Lyapunov function be $V_{a}=k_{a1}e_{a}^{T}e_{a}+\frac{1}{2}\dot{e}_{a}^{T}%
\dot{e}_{a}$, we can obtain that $e_{a}\rightarrow 0$ and $\dot{e}%
_{a}\rightarrow 0$ as $t\rightarrow \infty $. This concludes the proof. $%
\blacksquare $

\end{document}